\documentstyle[aps,eqsecnum,preprint,prd,epsfig]{revtex}
\begin{document}
\addtolength{\jot}{10pt}
\tighten
\newcommand{\spur}[1]{\not\! #1 \,}
\newcommand{\be}{\begin{equation}}
\newcommand{\ee}{\end{equation}}
\newcommand{\bea}{\begin{eqnarray}}
\newcommand{\eea}{\end{eqnarray}}
\newcommand{\ttp}{ {\tau \over \tau^\prime}}
\newcommand{\tp}{ {\tau^\prime \over \tau}}

\draft
\preprint{\vbox{\hbox{BARI-TH/98-290 \hfill}}}
\vskip 1cm
\title{\bf Universal $\tau_{1/2}(y)$ Isgur-Wise function \\
at the next-to-leading order  in QCD sum rules\\}
\vskip 0.5cm
\author{P.~Colangelo $^{1}$, F.~De~Fazio $^{1}$
and N.~Paver $^{2}$\footnote{Supported in part by MURST - Ministero della 
Ricerca Scientifica e Tecnologica.}
\\ }

\vskip 2.0cm
\address{$^{1}$ Istituto Nazionale di Fisica Nucleare, 
 Sezione di Bari,  Italy \\
$^{2}$ Dipartimento di Fisica Teorica, Universit\'a
di Trieste, Italy, and  \\
Istituto Nazionale di Fisica Nucleare, Sezione di Trieste, Italy\\}

\maketitle
\vskip 0.5cm
\begin{abstract}
We use QCD sum rules, in the framework of the Heavy Quark Effective 
Theory, to calculate the universal form factor $\tau_{1/2}(y)$ parameterizing 
the semileptonic transitions $B \to D_0 \ell \bar \nu$, 
$B \to D_1^* \ell \bar \nu$, where $D_0$ and  $D^*_1$ are the members 
of the excited charmed doublet with $J^P=(0^+_{1/2},1^+_{1/2})$. 
We include two-loop corrections in the perturbative contribution to the 
sum rule, and present a complete next-to-leading order result.
As a preliminary part of our analysis we also compute, up to order $\alpha_s$,
the leptonic constant $F^+$ of the doublet $D_0$, $D_1^*$.
Finally, we discuss the phenomenological implications of this calculation.

\end{abstract}

\vspace{1truecm}


\clearpage
\section{Introduction}

The application of the heavy quark flavor and spin symmetry, 
valid in QCD in the infinite heavy quark mass limit 
\cite{hqet_0,hqet_1}, together with the
heavy quark effective field theory (HQET) \cite{hqet_2,falk90,reviews}, 
has led to   
a dramatic progress towards a model-independent description of the 
spectroscopy and the decays of hadrons containing a single heavy quark $Q$
($Q=c,b$). 
An outstanding result of the theory concerns the description of the  
exclusive $B \to D \ell \bar\nu $ and $B \to D^* \ell \bar\nu$ semileptonic 
decays, in the limit $m_Q\to\infty$,  in terms of just one  
nonperturbative, universal form factor (the Isgur-Wise function $\xi$), 
normalized to unity at maximum momentum transfer to the lepton pair. Other 
distinctive examples are the relations between the beauty meson leptonic 
constants and the beauty meson semileptonic
transition amplitudes to light mesons at zero recoil, with the analogous 
charmed meson ones, obtained employing general dimensional scaling rules.
 
Corrections of order $\Lambda_{QCD}/m_Q$ (and higher powers) to the leading 
term can be systematically analyzed in 
HQET in terms of a reduced number of hadronic, universal parameters,
with a remarkable simplification of the analysis.
However,  
in the applications of HQET the effects of non-perturbative strong 
interactions  can be estimated only in the framework of some non-perturbative 
theoretical approach. In this regard, particularly fruitful has been the
application of sum rules \cite{shiflibro} formulated in the framework of 
HQET \cite{neubert_rev}. This method is genuinely field theoretical and 
based on first principles, and relates the hadronic observables to QCD 
fundamental parameters {\it via}
the Operator Product Expansion (OPE) of suitable 
Green's functions. Such an expansion  involves
perturbative contributions as well 
as non-perturbative quark and gluon vacuum condensates. 
In particular, $\alpha_s$ corrections to the coefficients of the OPE
can be computed order by order in perturbation theory,
and therefore they can be systematically taken into account.  

A critical aspect of the sum rule calculations in HQET
is represented by the size of non-leading terms,  such as the $1/m_Q$ 
corrections and the $\alpha_s$ corrections in the 
perturbative expansion of the OPE. For example, 
the predictions for the leptonic constants 
of $\bar q Q$ pseudoscalar mesons
are affected by considerably 
large next-to-leading corrections in $\alpha_s$ 
\cite{broad92,neubert92,bagan92}; also $\Lambda_{QCD}/m_Q$ corrections 
are non-negligible 
in the case of the $D$ meson,
an effect confirmed by lattice QCD analyses \cite{flynn97}. 

Conversely, in the HQET QCD sum rule calculation 
of the Isgur-Wise function, 
the next-to-leading order $\alpha_s$ corrections 
turn out to be small and well under control \cite{neubert93,bagan93}, 
and  the same is 
true for $\Lambda_{QCD}/m_Q$ corrections \cite{neubert93a}, 
specially near the zero recoil point 
where the normalization of the universal form factor is protected by the heavy 
quark symmetry. This has allowed a drastic reduction of the 
theoretical uncertainty in the determination of the CKM matrix element $V_{cb}$ 
\cite{neubert_peking}.

It is worth analyzing other cases analogous to the determination of the 
Isgur-Wise form factor  $\xi$,  and we present here a HQET sum rule 
calculation of the universal form factor governing the 
semileptonic $B$ meson decays into the $0^+, 1^+$ charmed excited states, up 
to next-to-leading order in $\alpha_s$ and to leading order in the heavy quark 
expansion $m_Q\to\infty$. These higher-lying charmed states correspond to the 
$L=1$ orbital excitations in the non-relativistic constituent quark model. 
Besides their theoretical relevance 
to HQET \cite{isgur91}, in particular to the aspects of the QCD sum rule 
calculation mentioned above, such $B \to D^{**}$ 
semileptonic transitions ($D^{**}$ is the generic $L=1$ charmed state)
have numerous additional points of physical interest. 

Indeed, in principle these decay modes may  
account for a sizeable fraction of semileptonic $B$-decays, and 
consequently they represent a well-defined set of corrections to the 
theoretical prediction that, in the limit 
$m_Q\to\infty$ and under the condition  $(m_b-m_c)/(m_b+m_c)\to 0$ (the 
so-called small-velocity limit),
the total semileptonic $B\to X_c$ decay rate should be saturated by 
the $B\to D$ and $B\to D^*$ modes \cite{hqet_1}. 
Moreover, the shape of the inclusive differential decay distribution 
in the lepton energy could reflect contributions from the 
$B\to D^{**}$  modes.  

Another important result, relevant both
to phenomenology and to the critical 
tests of HQET, is the relation of the $B\to D^{**}$ form factors 
at zero recoil to the slope of the $B\to D^{(*)}$ 
Isgur-Wise function, through the Bjorken sum rule \cite{bj90}. 
Of similar interest for HQET is the test of the upper bound  
on such universal form factors at zero recoil, involving the heavy meson 
``binding energy'' and the $D^{**}-D$ mass splittings, 
that is the analog of the 
``optical'' sum rule for dipole
scattering of light in atomic physics \cite{vol92,uraltsev}. 
 
Moreover, the $\Lambda_{QCD}/m_Q$ corrections can have a role for 
$B$-decay modes into excited charmed states, that mostly occur near the zero 
recoil point where the corresponding transition matrix elements vanish. The 
shape of the lepton energy spectrum near such kinematical 
point including the $1/m_Q$ corrections, that in HQET can be predicted in 
terms of the Isgur-Wise function and mesons mass splittings, 
represents an important test of the theory \cite{ligeti97}.     

Continuing with the aspects justifying the interest for $B \to D^{**}$, 
let us notice that
the investigation of the
semileptonic $B$ transitions to excited charm states
is an important preliminary study for the theoretical analysis of the 
production of such states in nonleptonic $B$ decays \cite{nonlepb},
as well as for the identification of additional decay modes 
(such as $D^{(*)} D^{(*)} \pi$) suitable for the investigation of CP violating 
effects at $B$ factories \cite{pene98}.

Finally, as a byproduct of the QCD sum rule calculation, theoretical 
predictions about the yet unobserved $D^{**}$ meson masses can be obtained, 
that are obviously interesting {\it per se}. 

In the following we present a complete next-to-leading order evaluation 
of the $B$-meson semileptonic transition to the scalar charmed state by 
QCD sum rules, at the leading order of $m_{b,c} \to \infty$.
In Sect. II we report the main aspects of the spectroscopy and 
decays of $L=1$
$\bar q Q$ mesons, together with the definition of the universal form factor 
$\tau_{1/2}(y)$. The various steps of the QCD sum rule determination of such 
form factor, within HQET, are collected in Sect. III and V-VII, together
with the analysis, in Sect. IV, of the leptonic constant $F^+$ of the
doublet $D_0,D^*_1$. In Sect. VIII the phenomenological implications of our 
calculation are presented, together with the conclusions.

\section{Positive Parity heavy-light mesons}

In the infinite heavy quark mass limit 
the spectrosopy of hadrons containing one heavy quark Q 
is greatly simplified, due to the decoupling of the heavy quark spin $\vec s_Q$
from the angular momentum of the light degrees of freedom
(quarks and gluons) $\vec s_\ell= \vec J - \vec s_Q$.
This allows a classification of such hadronic states
by $\vec J$ and  by $\vec s_\ell$,
so that hadrons corresponding to the same $s_\ell$ belong to degenerate 
doublets. 
In the case of $\bar q Q$ mesons,
the low-lying states with $s_\ell^P={1\over 2}^-$ correspond to the 
pseudoscalar $0^-$ and vector $1^-$ mesons
($B,B^*$; $D,D^*$), the $s$-wave states of the constituent quark model.
The four states corresponding
to orbital angular momentum $L=1$ 
can be classified in two doublets: 
$J^P=(0^+_{1/2},1^+_{1/2})$ and
$J^P=(1^+_{3/2},2^+_{3/2})$, which differ by the values 
$s_\ell^P={1\over 2}^+$ and $s_\ell^P={3 \over 2}^+$, respectively.
The states $0^+_{1/2}$ and $2^+_{3/2}$ correspond to the scalar and spin-two
mesons of the quark model; the relation between HQET and 
quark model $L=1$, $1^+$ states can easily be derived \cite{rosner86}.

The charmed $2^+_{3/2}$ state  has been experimentally observed
and denoted as the $D_2^*(2460)$ meson, with 
$m_{D_2^*}=2458.9\pm 2.0$ MeV, $\Gamma_{D_2^*}=23\pm 5$ MeV and
$m_{D_2^*}=2459\pm 4$ MeV, $\Gamma_{D_2^*}=25^{+8}_{-7}$ MeV for the neutral 
and charged states, respectively \cite{pdg}. 
The HQET state $1^+_{3/2}$ can be identified with $D_1(2420)$, with
$m_{D_1}=2422.2\pm 1.8$ MeV and 
$\Gamma_{D_1}=18.9^{+4.6}_{-3.5}$  MeV  \cite{pdg},
even though a $1^+_{1/2}$ component can be contained in such physical
state due to the mixing allowed for the finite value of the charm quark mass. 
\footnote{There is also experimental evidence of beauty
$s_\ell^P={3\over 2}^+$ states \cite{bstar}.}

Both the states $2^+_{3/2}$ and $1^+_{3/2}$ decay to hadrons by
$d-$wave transitions, which explains their narrow width; 
the strong coupling constant governing their two-body decays
can be determined using  experimental information \cite{falk92}.
On the other hand, the $s_\ell^P={1\over 2}^+$ doublet ($D_0, D^*_1$) 
has not been observed yet.
The strong decays of such states occur through $s$-wave transitions,
with expected larger
widths than in the case of the 
doublet ${3\over 2}^+$. Indeed, analyses of 
the coupling constant governing the two-body hadronic transitions
by QCD sum rules predict
$\Gamma(D_0^0 \to D^+ \pi^-) \simeq 180$ MeV and
$\Gamma(D_1^{*0} \to D^{*+} \pi^-) \simeq 165$ MeV \cite{colangelo95} . 
Estimates of the mixing 
angle $\alpha$ between $D^*_1$ and  $D_1$ give $\alpha \simeq 
16^0$  \cite{colangelo95,kilian92}.

The matrix elements of the semileptonic 
$B \to D_0 \ell \bar \nu $ and $B \to D_1^* \ell \bar \nu$  
transitions can be parameterized in terms of six form factors: 
\bea
{<D_0(v')|{\bar c}\gamma_\mu \gamma_5 b|B(v)> 
\over {\sqrt {m_B m_{D_0}}}  }
&=& g_+ (v+v^\prime)_\mu+
 g_- (v-v^\prime)_\mu \nonumber \\
{<D^*_1(v',\epsilon)|{\bar c}\gamma_\mu(1-\gamma_5)b|B(v)>
\over {\sqrt {m_B m_{D^*_1}}} }
&=& g_{V_1} \epsilon^*_\mu +\epsilon^* \cdot v \; [g_{V_2} v_\mu+
g_{V_3} v^\prime_\mu]
-i \; g_A
\epsilon_{\mu \alpha\beta \gamma} \epsilon^{*\alpha} v^\beta v^{\prime \gamma}
\;\;\;,  \label{full}
\eea
where $v$ and $v^\prime$ are four-velocities
and $\epsilon$ is the $D_1^*$ polarization vector. 
The form factors $g_i$ depend on the variable $y=v \cdot v^\prime$, which
is directly related to the momentum transfer to the lepton pair. 
In terms of such form factors the semileptonic differential decay rates can be 
expressed as
\bea
{d \Gamma \over dy}(B \to D_0 \ell \nu)&=&{G_F^2 |V_{cb}|^2 m_B^5 \over 48 
\pi^3}(y^2-1)^{3/2}\Big({m_{D_0} \over m_B}\Big)^3
 \Big[ \Big(1+{m_{D_0} \over m_B}\Big)g_+ -
\Big(1-{m_{D_0} \over m_B}\Big)g_- \Big]^2 \nonumber \\
{d \Gamma \over dy}(B \to D_1^* \ell \nu)&=&{G_F^2 |V_{cb}|^2 m_B^5 \over 48 
\pi^3}\sqrt{y^2-1}\Big({m_{D_1^*} \over m_B}\Big)^3
\Big\{ \Big[\Big(y-{m_{D_1^*} \over m_B}\Big)g_{V_1}+(y^2-1) \Big(g_{V_3}+
{m_{D_1^*} \over m_B}g_{V_2}\Big)\Big]^2
\nonumber \\&+&2\Big[1-2{m_{D_1^*} \over m_B}y+
\Big({m_{D_1^*} \over m_B}\Big)^2 \Big] [g_{V_1}^2+(y^2-1)g_{V_A}^2] \Big\}
\;\;\;. \label{rates}
\eea
 
The heavy quark spin symmetry allows to relate the form factors $g_i(y)$ in 
(\ref{full}) to a single function $\tau_{1/2}(y)$
\cite{isgur91} through short-distance coefficients, perturbatively calculable, 
which depend on the heavy quark masses $m_b,m_c$,
on $y$ and on a mass-scale $\mu$, and connect the QCD vector and axial vector 
currents to the HQET currents. At the next-to-leading 
logarithmic approximation in $\alpha_s$ and in the infinite heavy quark 
mass limit,
the relations between $g_i$ and $\tau_{1/2}$ are given by 
\bea
g_+(y) + g_-(y) &=& -2 \;\Big( C_1^5(\mu)+(y-1) C_2^5(\mu)\Big) 
\; \tau_{1/2}(y,\mu) \nonumber \\
g_+(y) - g_-(y) &=& 2\;\Big( C_1^5(\mu)-(y-1)C_3^5(\mu)\Big)
 \; \tau_{1/2}(y,\mu) \nonumber \\
g_{V_1}(y)&=&2(y-1) \;C_1(\mu) \; \tau_{1/2}(y,\mu) \nonumber \\
g_{V_2}(y)&=&-2\; C_2(\mu)\; \tau_{1/2}(y,\mu) \nonumber \\
g_{V_3}(y)&=&-2\;\Big(C_1(\mu)+C_3(\mu)\Big) \;\tau_{1/2}(y,\mu) \nonumber \\
g_A(y)&=&-2 \; C_1^5(\mu) \; \tau_{1/2}(y,\mu) \;. 
\label{rel_formf}
\eea
The $\mu$-dependence is the 
same for all the functions $C_i^{(5)}$, where $C_i$ refer to the vector current 
and $C_i^5$ to the axial one; therefore, one can extract such a dependence by
writing \cite{neubert_rev}
\be
C_i^{(5)}(\mu)={\hat C}_i^{(5)}(m_b,m_c,y) \; K_{hh}(y,\mu)
\label{nextcoef}
\ee
where $K_{hh}=[\alpha_s(\mu)]^{-a_{hh}(y)} \Big\{1-{\alpha_s(\mu) \over \pi} 
Z_{hh}(y) \Big\}$, with 
$a_{hh}={2 \over 9} \gamma(y)$, 
\be
\gamma(y)={4 \over 3}[yr(y)-1]
\label{andim}
\ee
and $r(y)={\ln (y+\sqrt{y^2-1}) \over \sqrt{y^2-1}}$,
is related to the velocity-dependent anomalous dimension of the heavy-heavy
$b\to c$ current in HQET \cite{falk90}.
The coefficient $Z_{hh}$,  derived in \cite{neubert_rev,neubert92,ji} 
has the expansion, for $n_f=3$:
\be
  Z_{\rm hh}(y) =\bigg({752\over 729} - {8\pi^2\over 81}\bigg)(y-1)
    - \bigg({368\over 1215} - {4\pi^2\over 135}\bigg)(y-1)^2
    + \ldots  \;\;\;. \nonumber
\ee
The next-to-leading-log expression of the coefficient functions 
${\hat C}_i^{(5)}$ can be found in \cite{neubert_rev}. At the leading-log 
approximation they simply read:
${\hat C}_1={\hat C}_1^5=\Big[{\alpha_s(m_b)\over \alpha_s(m_c)}\Big]^{a_I} 
[\alpha_s(m_c)]^{a_{hh}}$, 
with $a_I=-{6\over 25}$, the coefficients  $\hat C_2^{(5)}$ and
$\hat C_3^{(5)}$ being zero.

From (\ref{rel_formf}) and (\ref{nextcoef}) it follows that the 
product $K_{hh}(y,\mu) \tau_{1/2}(y,\mu)$ does not depend on $\mu$ , and 
therefore 
a renormalization-group invariant form factor can be defined
\be
\tau_{1/2}^{ren}(y)=K_{hh}(y,\mu) \tau_{1/2}(y,\mu)\;; \label{tauren}
\ee
hence, in (\ref{rel_formf}) one can substitute $C_i^{(5)}(\mu)$ with 
${\hat C}_i^{(5)}$ and $\tau_{1/2}(y,\mu)$ with $\tau_{1/2}^{ren}(y)$.

Analogous relations hold for the eight form factors parameterizing the matrix 
elements of $B \to D_1 \ell \bar \nu$ and
$B \to D^*_2 \ell \bar \nu$; in this case the heavy quark symmetry allows to 
relate them  to  another universal function $\tau_{3/2}(y)$
\cite{isgur91}. The main difference with respect to the Isgur-Wise form factor 
$\xi(y)$ is that one cannot invoke symmetry arguments to predict the 
normalization of both $\tau_{1/2}(y)$ and $\tau_{3/2}(y)$, and therefore a 
calculation of the form factors in the whole kinematical range is required.
For $B \to (D_0, D^*_1) \ell \bar \nu$
the physical range for the variable $y$ is restricted between
$y=1$ and $y=1.309-1.326$, taking into account 
the values for the mass of $D_0$, $D^*_1$
($m_{D_0,D^*_1}=2.40-2.45$ GeV). Consequently,
 for practical purposes one can 
adopt a linear parameterization of the form factor $\tau_{1/2}(y)$: 
$\tau_{1/2}(y) \simeq \tau_{1/2}(1) (1- \rho_{1/2}^2 (y-1))$. This 
parameterization allows an easier comparison between the predictions of 
different approaches.

A determination of $\tau_{1/2}(y)$ by QCD sum rules at ${\cal O}(\alpha_s=0)$
was carried out in \cite{cola92}, starting from finite values of the beauty and 
charm quark masses and the performing the limit $m_b,m_c \to \infty$. 
The obtained result can be summarized by the values 
$\tau_{1/2}(1) \simeq 0.25$ and $\rho^2_{1/2} \simeq 0.4$, 
which imply quite small values 
of the branching ratios  of $B\to (D_0,D^*_1) \ell \bar \nu$.
 
Other determinations of $\tau_{1/2}(y)$
have appeared in the literature, 
employing various versions of the constituent quark model
\cite{qmodel,wambach,veseli,morenas,deandrea}. 
The results range in a quite large interval,
$\tau_{1/2}(1)=0.06-0.40$ and
$\rho^2_{1/2}=0.7-1.0$, and critically depend on the peculiar
features of the models employed in the numerical calculation.

As for $\tau_{3/2}(y)$, a QCD sum rule analysis to the leading order in
$\alpha_s=0$ \cite{cola92} gives
$\tau_{3/2}(1)\simeq 0.28$ and $\rho^2_{3/2}\simeq 0.9$.
Quark model results, on the other hand, give predictions in the range
$\tau_{3/2}(1)\simeq 0.31-0.66$ and $\rho^2_{3/2}\simeq 1.4-2.8$
\cite{qmodel,wambach,veseli,morenas,deandrea}. 
We do not consider here the problem of the role of radiative 
corrections to the function $\tau_{3/2}$, but limit our analysis to the case of 
$\tau_{1/2}$, for which a number of interesting information can be worked out.

In the next Section we briefly outline the basic points of the QCD sum rule 
method, as needed for the extension of the calculation of \cite{cola92}
to the next-to-leading order in $\alpha_s$.

\section{Form factor $\tau_{1/2}$ from QCD sum rules in HQET}

Following \cite{neubert_rev},
the determination of the universal function $\tau_{1/2}(y)$ by QCD sum rules
in HQET is based on the analysis of the three-point correlator 
\cite{cola92}
\begin{eqnarray}
\Pi_\mu(\omega, \omega^\prime, y) &=& i^2 \int dx \; dz e^{i(k' x-k z)}
<0|T[ J_s^{v'}(x), {\tilde A}_\mu(0), J_5^{v}(z)^\dagger]|0> \nonumber \\
&=& i (v-v^\prime)_\mu \Pi(\omega, \omega^\prime, y)\;\;\;,
\label{threep}
\end{eqnarray}
where
${\tilde A}_\mu= \bar h_{Q^\prime}^{v'} \gamma_\mu \gamma_5 h_Q^v$ is the 
$b \to c$ weak axial current, and
$J_s^{v'}=\bar q h_{Q^\prime}^{v\prime}$ and 
$J_5^{v}=\bar q i \gamma_5 h_Q^v$  
represent local  interpolating currents of the
 scalar ($D_0$) and pseudoscalar ($B$) mesons in eq.(\ref{full})
represented in terms of HQET $h_Q^v$ fields and light quark $q$ fields. 
\footnote{For an extensive  discussion of $\bar q Q$ meson interpolating 
currents in HQET see, e.g.,  \cite{dai}.}
The variables $k, k^\prime$ are "residual" momenta, obtained by
the expansion of the heavy meson momenta in terms of the four-velocities:
$P=m_Q v+ k$, $P^\prime=m_{Q^\prime} v^\prime+ k^\prime$; 
they are ${\cal O}(\Lambda_{QCD})$, and remain finite in 
the heavy quark limit. 

Using the analyticity of $\Pi(\omega, \omega^\prime, y)$ in the variables
$\omega=2 v\cdot k$ and $\omega^\prime=2 v^\prime\cdot k^\prime$ at fixed $y$,
one can represent the correlator 
(\ref{threep}) by a double dispersion 
relation of the form
\begin{equation}
\Pi(\omega, \omega^\prime, y) = \int d \nu d \nu^\prime
{\rho(\nu, \nu^\prime, y) \over (\nu - \omega - i \epsilon) 
(\nu^\prime - \omega^\prime - i \epsilon) } \;\; \;, \label{correlatore}
\end{equation}
apart from possible subtraction terms. The correlator 
$\Pi(\omega, \omega^\prime, y)$
receives contributions from poles located at positive real
values of $\omega$ and $\omega^\prime$, corresponding to the physical 
single particle hadronic states 
in the spectral function $\rho(\nu, \nu^\prime, y)$.
The lowest-lying contribution is represented by 
the $0^-, 0^+$ $\bar q Q$ and  $\bar q Q^\prime$ states, i.e. 
$B$ and $D_0$.
This contribution introduces 
the form factor $\tau_{1/2}$ through the relation:
\begin{equation}
\Pi_{pole}(\omega, \omega^\prime, y) = 
{-2 \tau_{1/2}(y, \mu)  F(\mu)  F^+(\mu)  \over 
(2 \bar \Lambda  - \omega - i \epsilon) 
(2 \bar \Lambda^+ - \omega^\prime - i \epsilon) } \;, \label{pole}
\end{equation}
where $\mu$ is a renormalization scale and $F(\mu)$, $F^+(\mu)$ 
are the
couplings of the pseudoscalar and scalar interpolating 
currents to the  $0^-$ and $ 0^+$  states, respectively,
 in the heavy quark theory:
\begin{eqnarray} 
<0| J_5^v |B(v) > &=& F(\mu) \label{f} \\
<0| J_s^v |B_0(v) > &=& F^+(\mu) \label{fhat} \;\;\;.
\end{eqnarray}
$F(\mu)$ and $F^+(\mu)$ are scale-dependent low energy HQET parameters, 
which do not depend on the heavy quark mass $m_Q$ or $m_Q^\prime$.
In particular, $F(\mu)$ is related to the $B$-meson leptonic decay constant 
$f_B$. 
The mass parameters $\bar \Lambda$ and $\bar \Lambda^+$ identify the position 
of the poles in $\omega$ and $\omega^\prime$, 
and can be interpreted as binding energies of the $0^-$ and 
$0^+$ states:
$\bar \Lambda=M_B - m_b$, $\bar \Lambda^+=M_{D_0} - m_c$.

The higher states contributions to $\rho(\nu, \nu^\prime,y)$ can be
taken into account by a QCD continuum starting at some thresholds $\nu_c$ and 
$\nu^\prime_c$,  and  are modeled by the asymptotic freedom, 
perturbative spectral 
function $\rho^{pert}(\nu, \nu^\prime,y)$ according to the quark-hadron duality 
assumption. Here, $\rho^{pert}$ is the absorptive part of the perturbative 
quark-triangle diagrams, with two heavy quark lines corresponding to the weak 
$b \to c$ vertex and one light quark line connecting the heavy meson 
interpolating current vertices in (\ref{threep}). At the next-to-leading 
order in $\alpha_s$, all possible internal gluon lines in such triangle 
diagrams must be considered.

Therefore, for the dispersive representation (\ref{correlatore}) 
in terms of hadronic 
intermediate states one assumes the ansatz
\be
\Pi(\omega, \omega^\prime, y)= \Pi_{pole}(\omega, \omega^\prime, y) +
\Pi_{continuum}(\omega, \omega^\prime, y) \label{ansatz}
\ee
where, for simplicity,
the dependence of the continuum contribution on the thresholds
$\nu_c$ and $\nu^\prime_c$ has been omitted.

The correlator
$\Pi(\omega, \omega^\prime, y)$ can be expressed in QCD in the 
Euclidean region, 
i.e. for large  negative values of $\omega$ and $\omega^\prime$, in terms of 
perturbative and nonperturbative contributions:
\begin{equation}
\Pi(\omega, \omega^\prime, y) = \int d \nu d \nu^\prime
{\rho^{pert}(\nu, \nu^\prime, y) \over (\nu - \omega - i \epsilon) 
(\nu^\prime - \omega^\prime - i \epsilon) }  + 
\Pi^{np}(\omega, \omega^\prime, y)
\;\;\;. \label{qcd}
\end{equation}
In (\ref{qcd})
$\Pi^{np}$ represents the series of power corrections in  the "small"
${1 \over \omega}$  and  ${1 \over \omega^\prime}$ variables,
determined by quark and gluon vacuum condensates ordered by increasing 
dimension. These "universal" QCD parameters account for general properties of 
the nonperturbative strong interactions, for which asymptotic freedom
cannot be applied. The lowest dimensional ones can be obtained from
independent theoretical sources, or fitted from other applications of QCD sum 
rules to cases where the hadronic dispersive contribution is particularly well 
known. In practice, since the higher dimensional condensates are not known, one 
truncates the power series and a posteriori verifies the validity of such an 
approximation. In our application we shall include the dimension three quark 
condensate and the dimension five quark-gluon mixed condensate, which are known 
rather reliably; we neglect the contribution of the gluon condensate,
which always turns out to be numerically small in the analysis of
heavy-light meson systems.
The calculation  of $\rho^{pert}$ and $\Pi^{np}$ at the next-to-leading
order $\alpha_s$ is described in the sequel. 

The QCD sum rule 
for $\tau_{1/2}$ is finally obtained by imposing that the two representations 
of $\Pi(\omega, \omega^\prime, y)$, namely the QCD representation (\ref{qcd}) 
and the pole-plus-continuum ansatz (\ref{ansatz}), match in a suitable range of 
Euclidean values of $\omega$ and $\omega^\prime$. 

A double Borel transform  in the variables 
$\omega$ and $\omega^\prime$
\begin{equation}
{1 \over \tau} \hat{\cal B}_\tau^{(\omega)}=  lim \; 
{\omega^n \over (n-1)!} \big( - {d \over d \omega} \big)^n 
\hskip 1cm (n\to \infty, \omega \to -\infty, \tau=- {\omega \over n} \; fixed )
\end{equation}
(and similar for $\hat{\cal B}_{\tau^\prime}$)
is applied to "optimize" the sum rule.
As a matter of fact, this operation has two effects. The first one consists in
factorially improving the convergence 
of the nonperturbative series, justifying the truncation procedure;
the second effect is to enhance the role of the lowest-lying meson states 
while minimizing 
that of the model for the hadron continuum.  The 
{\it a priori} undetermined mass parameters $\tau$ and $\tau^\prime$ must
be chosen in a suitable range of values, in the present application
expected to be 
of the order of the typical hadronic 
mass scale $(\ge 1\; GeV)$, where the optimization is verified and,
in addition, the prediction turns out to be 
reasonably stable.
After the Borel transformation, possible subtraction terms are eliminated and
 eq.(\ref{correlatore}) can be rewritten as
\be
\hat \Pi (\tau, \tau^\prime, y)=\int d \nu d \nu^\prime \; e^{-\Big({\nu \over 
\tau}+{\nu^\prime \over \tau^\prime}\Big)} \rho(\nu, \nu^\prime, y)\;. 
\label{bor}
\ee

Eq. (\ref{pole}) shows that the preliminary
evaluation of the constants $F(\mu)$ and $F^+(\mu)$ is necessary to exploit 
the sum rule for the determination of $\tau_{1/2}$. 
This calculation is discussed in the next Section.

\section{Determination of $F^+(\mu)$ and $\bar \Lambda^+$}

The QCD sum rule determination of  $F^+(\mu)$ can be done by analyzing 
the correlator
\be
\Psi(\omega^\prime)=i \int d^4x \; e^{ik^\prime\cdot x}
<0|T[J_s^{v^\prime}(x) J_s^{v^\prime}(0)^\dagger]|0> 
\;\;\;, \label{corr2pt} 
\ee
\noindent 
whose dispersive representation takes contribution from the $0^+$ 
$\bar q Q$ pole
\be
\Psi(\omega^\prime)={[F^+(\mu)]^2 \over 2 \bar \Lambda^+-
\omega^\prime-i \epsilon}+ {1 \over 
\pi} \int_{\nu^\prime_c}^{+\infty} d \nu^\prime {Im\Psi(\nu^\prime) \over 
\nu^\prime-\omega^\prime -i \epsilon}+ \;subtr.\;, \label{disp}
\ee
$\nu^\prime_c$ being  
the effective threshold separating the contribution of the first resonance 
from the continuum.

It is straightforward to derive, at 
the next-to-leading order in $\alpha_s$, the contributions to the perturbative 
and non-perturbative parts of $\Psi(\omega^\prime)$. 
In the $\overline{MS}$ scheme one obtains
\be
{1\over \pi}Im \Psi^{pert}(\omega^\prime)=
{3 \omega^{^\prime 2} \over 8 \pi^2} \Theta (\omega^\prime) \Big[1+
{2 \alpha_s \over \pi} \Big(\ln\Big({\mu \over \omega^\prime}\Big)
+{17 \over 6} + {2 
\pi^2 \over 9}\Big) \Big] \label{impi}
\ee
\noindent and, considering non-perturbative vacuum condensates 
up to dimension five,
\bea
\Psi^{<\bar q q>}(\omega^\prime)&=&-{<\bar q q>(\mu) \over \omega^\prime} 
\Big[1+{2 \alpha_s \over \pi} \Big] \label{fhatdim3}
\\
\Psi^{<GG>}(\omega^\prime)&=&0 \label{fhatdim4}
\\
\Psi^{<\bar q G q>}(\omega^\prime)&=& {m_0^2 <\bar q q> \over 2 
\omega^{^\prime3}} 
\label{fhatdim5}
\eea
where the 
relation $<\bar q g_s \sigma \cdot G q>=m_0^2 <\bar q q>$ has been used
($m_0^2=0.8 \pm 0.2 \; GeV^2$ \cite{shiflibro}).
Consistently with the first order in $\alpha_s$ considered here, we neglect 
perturbative corrections to the coefficients of the higher-dimensional 
condensates in (\ref{fhatdim4}) and (\ref{fhatdim5}). 
The scale-dependence of the quark condensate is 
\be 
<\bar q q >(\mu)=<\bar q q>(\mu_0)\; \Big({\alpha_s(\mu) \over \alpha_s(\mu_0
)}\Big)^{-d} \;, \label{condrun}
\ee
\noindent where $\displaystyle{d={12 \over 33 -2n_f}}$, 
$n_f$ is the number of "active" quarks and
\be
\alpha_s(\mu)={12\pi \over 33 -2n_f}\; {1 \over \ln\Big({\mu^2 \over 
\Lambda_{\overline{MS}}^2}\Big)} \Bigg[1-6{153-19n_f \over (33 -2n_f)^2}
 {\ln \Big(\ln\Big({\mu^2 \over \Lambda_{\overline{MS}}^2}\Big)\Big) \over 
\ln\Big({\mu^2 \over \Lambda_{\overline{MS}}^2}\Big)} \Bigg] \;\;.
\ee 
The numerical value we use for the quark condensate at $\mu_0=1\;GeV$ 
is $<\bar q q>(\mu_0)=(-240 \;MeV)^3$. 

The Borel improved sum rule for $F^+$ reads
\bea
[F^+(\mu)]^2 e^{-2{ \bar \Lambda^+ \over \tau^\prime}}&=& {3 \over 8 \pi^2} 
\int_0^{\nu^\prime_c} 
d \nu^\prime \; \nu^{\prime 2} \Big[1 + {2 \alpha_s \over \pi} 
\Big(\ln\Big({\mu \over \nu^\prime}\Big) +{17 \over 6} + 
{2 \pi^2 \over 9}\Big) \Big]
  e^{-{\nu^\prime \over \tau^\prime}}\nonumber \\
&+& <\bar q q>(\mu) \Big[1+{2 \alpha_s \over \pi}-
{m_0^2 \over 4 \tau^{\prime2}} \Big] 
\;.\label{fhatsumrule}
\eea      
Eq.(\ref{fhatsumrule}) shows that, for the renormalization scale $\mu$,
hence for the argument of $\alpha_s$, of  the order of a typical 
strong interaction scale, $\simeq 1 - 2 \; GeV$,
the next-to-leading contribution to the perturbative part of the sum rule
for $F^+$ is large, similar to the situation met in the case of $F$ 
\cite{broad92,neubert92,bagan92}.

A $\mu$-independent constant $F^+_{ren}$ can be 
defined, using eq. (\ref{fhatsumrule}) and the relation 
between $F^+(\mu)$ and the matrix element of the  scalar current in
full QCD, in the same way one defines a $\mu$-independent leptonic constant
$F_{ren}$ for the $s_\ell^P={1\over 2}^-$ doublet
\cite{neubert93}:
\be
F^+_{ren}=[\alpha_s(\mu)]^{d \over 2} \Big[
1-{\alpha_s(\mu) \over \pi} \; Z] \;F^+(\mu)
\ee
where, in the  $\overline{MS}$ scheme, $\displaystyle
{Z=3{153-19n_f \over (33-2n_f)^2}-{381-30n_f+28\pi^2 \over 36 (33-2 n_f)}}$. 

\begin{figure}
\begin{center}
\mbox{%
\psfig{file=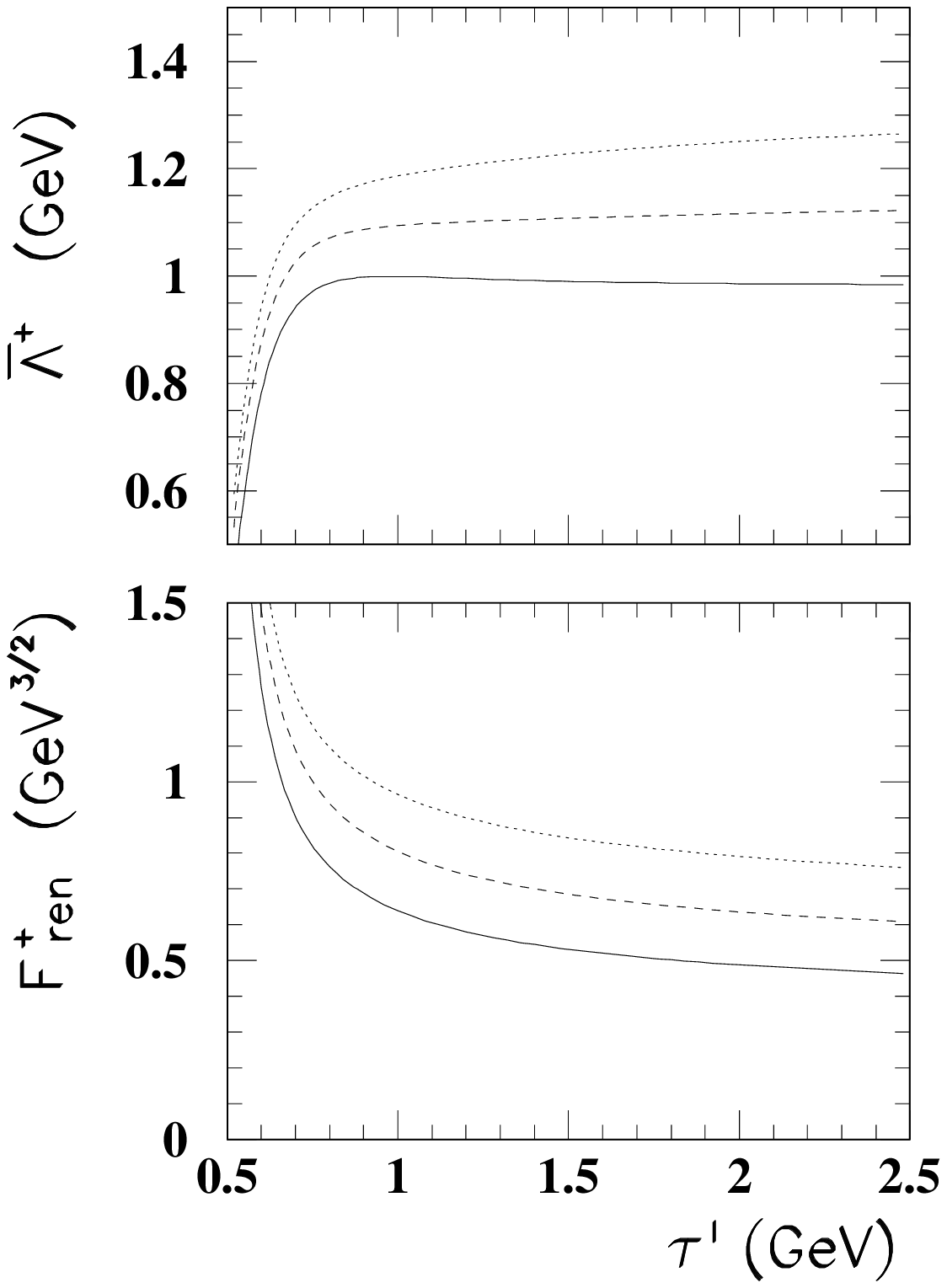,width=12cm}}
\end{center}
\caption{ } \label{fig_fpiu}
\end{figure}

From the above equations, a determination of $F^+(\mu)$ and 
$F^+_{ren}$ can be obtained, together with the parameter $\bar \Lambda^+$. 
The latter 
quantity, for example, can be evaluated 
by considering the logarithmic derivative 
of eq. (\ref{fhatsumrule}) with respect to the Borel parameter $\tau^\prime$. 
With $\tau^\prime$ in the range $(1-2.5)\;GeV$ and the 
threshold $\nu^\prime_c$ in the 
range $(2-3)\;GeV$, and choosing
$\Lambda_{QCD}=380 \; MeV$, we obtain for 
$\bar \Lambda^+$ and $F^+_{ren}$ the curves depicted in fig.\ref{fig_fpiu}. 
The corresponding predictions are:
\be
\bar \Lambda^+= 1.0 \pm 0.1 \; GeV \hskip 2 cm 
F^+_{ren}=0.7 \pm 0.2 \;GeV^{3\over 2}\;. \label{fpiun}
\ee

The expressions relevant to sum rule for the constant $F(\mu)$ 
in (\ref{f}) referring to the $(0^-,1^-)$ 
heavy meson doublet, are given by eq. (\ref{impi}) for the perturbative part, 
and by reversing the signs of (\ref{fhatdim3}-\ref{fhatdim5})
for the vacuum condensate contributions. 
An extensive analysis of this quantity can be found in \cite{neubert92}.
We only repeat here the numerical calculation of \cite{neubert92}, 
using the same input parameters 
adopted for $F^+$, and choosing the continuum threshold in the range 
$\nu_c=2-3$ GeV. The result is
\be
\bar \Lambda= 0.5 \pm 0.1 \; GeV \hskip 2 cm  
F_{ren}=0.45 \pm 0.05 \; GeV^{3\over 2}\;\;\;. \label{fres}
\ee
The difference $\Delta= \bar \Lambda^+ - \bar \Lambda$
corresponds to the difference between 
$m_{\bar D}$ and  $m_{\bar D_0}$, where
$\bar D$ and  $\bar D_0$ are the spin averaged states of the 
${1\over 2}^-$ and ${1\over 2}^+$ doublets. Our central value
$\Delta=0.5 \; GeV$ allows us to predict $m_{\bar D_0}\simeq 2.45 \; GeV$ 
with an uncertainty of about $0.15$ GeV.

It is worth reminding that determinations of
the leptonic constant $F^+$ at the order $\alpha_s=0$ by QCD sum rules  gave 
the result: $F^+=0.46 \pm 0.06$ GeV$^{3\over 2}$ \cite{cola92}
and $F^+=0.40 \pm 0.04$ GeV$^{3\over 2}$ and ${\bar \Lambda}^+=1.05 \pm 0.5 
\; GeV$ or ${\bar \Lambda}^+=0.90 \pm 0.10 
\; GeV$ \cite{dai,dai97} 
depending on the choice of the interpolating currents.
The difference with respect to the values in
(\ref{fpiun}) is the effect of sizeable  
radiative corrections. However, as far as the
determination of $\tau_{1/2}$ is concerned, 
since radiative corrections affect both  
the three-point correlator, and the two-point
functions determining the leptonic constants, it is still
possible that a partial
compensation occurs in the ratio determining the form factor;
we shall see in the following that this is, indeed, the case.

Determinations of $F^+$ by quark models 
\cite{oliver97} give results in agreement with 
(\ref{fpiun}) when  relativistic models  \cite{rqmod} are employed:
$F^+ \simeq 0.6 - 0.7 \; GeV^{3/2}$. Conversely,
lower values are obtained: $F^+ \simeq 0.235 \; GeV^{3/2}$ 
using  non relativistic models \cite{nrqmod}.

The values in (\ref{fpiun}) and (\ref{fres}), or the equations
corresponding to the respective sum rules,
can be used as an input  in (\ref{pole}) to determine $\tau_{1/2}$.

\section{${\cal O}(\alpha_s)$ corrections to the sum rule for $\tau_{1/2}$}

\begin{figure}
\begin{center}
\mbox{%
\psfig{file=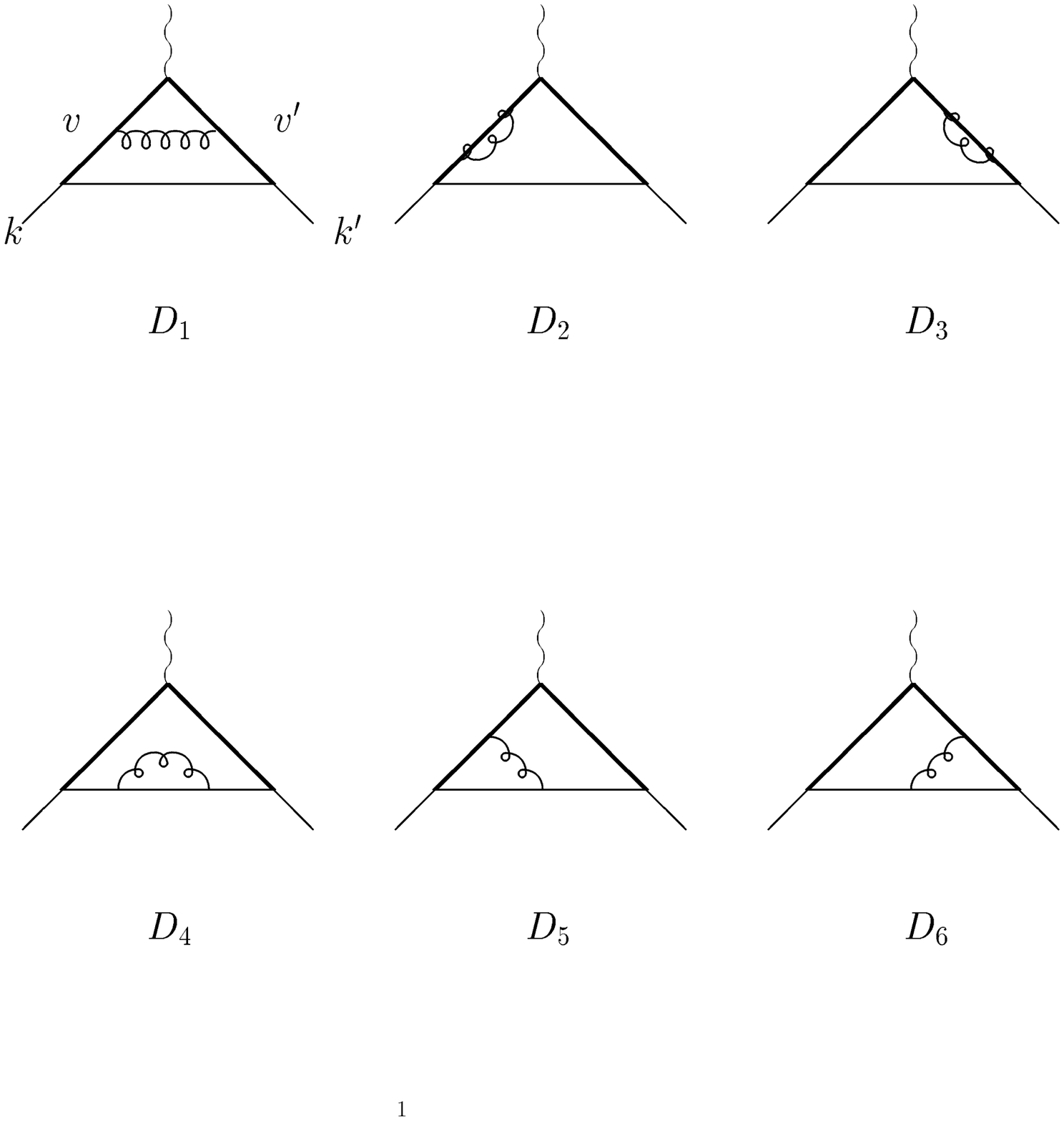,width=12cm}}
\vskip -2cm
\end{center}
\caption{ } \label{diagrams_pert}
\end{figure}

In order to calculate ${\cal O}(\alpha_s)$ corrections to the
perturbative part of the sum rule for 
$\tau_{1/2}(y)$, one has to compute the two-loop diagrams 
depicted in fig.\ref{diagrams_pert}.
Also the non perturbative term proportional to the 
quark condensate  receives ${\cal O}(\alpha_s)$ corrections, as 
discussed in Section VI. On the other hand, consistently with the order in 
$g_s$ considered here and with the previous estimates of $F^+$ and
$F$, radiative corrections 
to the contributions of higher dimensional 
condensates will not be included.

We start from the calculation of the perturbative part.
As shown in \cite{neubert93}, it is useful to directly deal with the 
double-Borel transformed expressions of the integrals 
corresponding to the various 
diagrams, a procedure which considerably simplifies the resulting calculations.
This is the strategy we follow here, keeping different values of the
Borel parameters $\tau$ and $\tau^\prime$ in (\ref{bor}). 
Adopting the standard dimensional regularization procedure,
we compute the diagrams 
in $D$ space-time dimensions, using the Feynman rules of HQET, and then 
we consider the expansion for $\epsilon=(D-4)/2 \to 0^+$.

At the order $\alpha_s=0$ the expression for the Borel-transformed correlator
(\ref{threep}) in the variables $\omega, \omega^\prime$
is given by
\be
{\hat D}_0=  i (v-v^\prime)_\mu (1- {\tau \over \tau^\prime}) 
{4 N_c \over (4 \pi)^{D/2} } { \Gamma(D/2) \;\; \tau^{D-1}
\over [V^2(\tau/\tau^\prime)]^{D/2} }\;,\label{d0}
\ee
where $N_c$ is the number of colours and $V^2(u)=u^2+2uy+1$;
$\tau$ and $\tau^\prime$ are the Borel variables related to 
$\omega$ and $\omega^\prime$, respectively.
Eq. (\ref{d0}) shows that one has to perform the calculation with $\tau 
\neq \tau^\prime$ from the very beginning;
a different situation is 
met in the case  of the Isgur-Wise form factor $\xi(y)$, where the choice 
$\tau=\tau^\prime$ is allowed by the symmetry of the three-point correlator,
with a remarkable simplification of the analysis.
The requirement of keeping different values of the Borel variables 
represents the main technical difficulty in this calculation.

In the following we give the results for the various diagrams in 
fig.\ref{diagrams_pert}, together with 
few details concerning 
the calculation; some useful formulae are
collected in the Appendix. The overall computational strategy follows 
that adopted in \cite{neubert93}
for the calculation of the $\alpha_s$ corrections to 
the Isgur-Wise function $\xi(y)$, 
and we refer to this paper for further details.

\subsection{Diagrams $D_1$, $D_2$, $D_3$}
Let us first consider the diagrams $D_1$, $D_2$ and $D_3$, 
where the gluon has both vertices on the heavy quark  lines. 
Applying HQET Feynman rules we obtain for the the diagram $D_1$:
\bea
D_1&=& 16 i N_c g_s^2 C_F y Tr\Big[\gamma_5  {1+ \spur{v} \over 2} \gamma_\mu 
\gamma_5 {1+ \spur{v}^\prime \over 2} \gamma^\alpha \Big] 
\;\;\;\label{d1}
\\ 
\int {d^D s \over (2\pi)^D} {d^D t \over (2\pi)^D}&& 
{s_\alpha \over (\omega+2 v \cdot s)(\omega+2 v \cdot t)
(\omega^\prime+2 v^\prime \cdot s)(\omega^\prime+2 v^\prime \cdot t)s^2(s-t)^2}
\nonumber
\eea
\noindent
and, after double-Borel transform in the variables $\omega, \omega^\prime$,
\be
{\hat D}_{1}={y \over D-4} \ttp Z_\mu (1-\ttp) \Bigg\{ \int_0^1 {du \over 
\Big[ V^2\Big(u\ttp\Big) \Big]^{D/2-1}}+ \Big(\tp\Big)^{D-2} 
\int_0^1 {du \over \Big[ V^2\Big(u\tp\Big) \Big]^{D/2-1}} \Bigg\}
 \label{d1hat}
\ee
\noindent where 
\be
Z_\mu=16 i N_c g_s^2 C_F {\tau^{2D-5} \over (4\pi)^D} 
\Gamma \Big( {D \over 2}-1 \Big) \Gamma \Big({D \over 2} \Big) 
{(v-v^\prime)_\mu \over \Big[ V^2\Big(\ttp\Big) \Big]^{D/2} } \label{z}  
\ee
and $C_F={N_c^2-1 \over 2 N_c}$.
By a similar calculation,   after borelization we obtain for the 
diagrams $D_2$ and $D_3$:
\be
\hat D_2=-Z_\mu \Big(1-\ttp\Big){1 \over (D-3)(D-4)}
\;\;\;  \label{d2hat} 
\ee
\be
\hat D_3=-Z_\mu \Big(1-\ttp\Big){1 \over (D-3)(D-4)}\Big(\tp\Big)^{D-4}
\;\;\; . \label{d3hat} 
\ee
\noindent
Expanding (\ref{d1hat})-(\ref{d3hat}) around $\epsilon=0$,
 and summing up the contributions of 
$\hat D_1-\hat D_3$, we obtain:
\bea
\sum_{i=1}^3\hat D_i={Z_\mu \over 2} \Big(1-\ttp\Big) 
&\Bigg\{&{1 \over \epsilon}
\Big[yr(y)-2\Big] \label{d123}\\
&+& y\Big[h\Big(y,\tp\Big)+h\Big(y,\ttp\Big)\Big]+y\;\ln{V^2\Big(\tp\Big)}r(y)
-y\sqrt{y^2-1}r^2(y) \nonumber \\
&-& 2y \sqrt{y^2-1}F_1\Big(\tp\Big)F_1\Big(\ttp\Big)+4-\ln\Big(\tp\Big)^2 
 \; + {\cal O}(\epsilon) \Bigg\} \nonumber
\eea
\noindent where 
$r(y)={\ln(y_+) \over \sqrt{y^2-1}}$ is the function already met in the 
expression of the Wilson coefficient in eq.(\ref{rel_formf}), and
the variable $y_\pm$ is defined as $y_\pm=y\pm \sqrt{y^2-1}$;
the function $h(y,z)$ is 
\be
h(y,z)={1 \over \sqrt{y^2-1}}\Big[-L_2(1-y_-^2)+L_2\Big({y_+(1-y_-^2) \over
y_+ +z}\Big) \Big] \;\;\;
\ee
with $L_2$ the dilogarithm 
function: $L_2(x)=-\int_0^x dt {\ln(1-t) \over t}$. 
The function $F_1(z)$ in (\ref{d123}) is reported in  the Appendix. 

\subsection{ Diagram $D_4$}
The expression of the diagram $D_4$, involving the light quark self-energy, 
is
\bea
D_{4}&=&4iN_c  g_s^2 C_F (2-D) Tr\Big[ \gamma_5 {1+ \spur{v} \over 2} 
\gamma_\mu 
\gamma_5 {1+ \spur{v}^\prime \over 2} \gamma^\alpha \gamma^\beta \gamma^\gamma
 \Big] \label{d4}\\
& & \int {d^D s \over (2\pi)^D} {d^D t \over (2\pi)^D} 
{s_\alpha t_\beta s_\gamma \over (\omega+2 v \cdot s)
(\omega^\prime+2 v^\prime \cdot s)(s^2)^2t^2(s-t)^2}\;.
\nonumber \eea
Integration over $t$ is straightforward. The double-Borel transformed
$\hat D_4$ is immediately obtained: 
\be
\hat D_4={Z_\mu \over 2}\Big(1-\ttp\Big){1 \over D-4}{1 \over 
\Big[V^2\Big(\ttp\Big)\Big]^{D/2-2}} \label{d4hat}
\ee
\noindent which, after expanding around  $\epsilon=0$, becomes
\be
\hat D_4={Z_\mu \over 2}\Big(1-\ttp\Big)\Bigg[{1 \over 2 \epsilon}-{1 \over 2}
\ln{V^2\Big(\ttp\Big)} +{\cal O}(\epsilon) \Bigg] \;. \label{d4exp}
\ee

\subsection{ Diagrams $D_5, D_6$}
The most difficult diagrams to compute are $D_5$ and $D_6$, 
where the gluon has one vertex on the light quark line.
Starting from the expression of $D_5$
\bea
D_5&=&8i N_c g_s^2 C_F Tr\Big[\gamma_5  {1+ \spur{v} \over 2} \gamma_\mu 
\gamma_5 {1+ \spur{v}^\prime \over 2} \gamma^\alpha  \spur{v} 
\gamma^\beta\Big] \label{d5}\\
\int {d^D s \over (2\pi)^D} {d^D t \over (2\pi)^D} 
&&{s_\alpha t_\beta  \over (\omega+2 v \cdot s) (\omega+2 v \cdot t)
(\omega^\prime+2 v^\prime \cdot s)s^2 t^2(s-t)^2} 
\nonumber \eea
\noindent and using the identity 
\be
Tr\Big[\gamma_5  {1+ \spur{v} \over 2} \gamma_\mu
\gamma_5 {1+ \spur{v}^\prime \over 2} \gamma^\alpha  \spur{v}
\gamma^\beta\Big]=Tr\Bigg[\gamma_5  {1+ \spur{v} \over 2} \gamma_\mu
\gamma_5 {1+ \spur{v}^\prime \over 2} \Big[ g^{\alpha \beta}+{1 \over 2}[
\gamma^\alpha,\gamma^\beta]+2v^\beta \gamma^\alpha \Big] \Bigg]
\ee one can 
write: $D_5=D_5^{(1)}+D_5^{(2)}+D_5^{(3)}$, in correspondence
to the above three terms. The first one, obtained by using
$s\cdot t={1\over2}[s^2+t^2-(s-t)^2]$, is given by
\bea
\hat D_5^{(1)}={Z_\mu \over 2}{1 \over D-2} &\Bigg\{&\Big(\tp \Big)^{D-2}
V^2\Big(\ttp\Big)\int_0^1{ du \over \Big[V^2\Big(u\tp\Big) \Big]^{D/2-1}} 
- {1 \over D-3}V^2\Big(\ttp\Big) \nonumber \\
&-& \Big(\tp \Big)^{D-2}\Big[V^2\Big(\ttp\Big) \Big]^{D/2} 
\int_0^1du { (1-u)^{2-D} \over \Big[V^2\Big(u\tp\Big) \Big]^{D/2-1}} \Bigg\} \;
\;,
\label{d5hat} 
\eea
where it is worth noticing the nontrivial dependence on the Borel parameters 
$\tau$ and $\tau^\prime$. 
Eq. (\ref{d5hat}) can be simplified by integrating 
by parts the last integral
\bea
\int_0^1du { (1-u)^{2-D} \over \Big[V^2\Big(u\tp\Big) \Big]^{D/2-1}}=
&-&{1 \over D-3}+{D-2 \over D-3} \tp \Big(y+\tp \Big) 
\int_0^1du { (1-u)^{3-D} \over \Big[V^2\Big(u\tp\Big) \Big]^{D/2}}
\nonumber \\
&-&{D-2 \over D-3} \Big(\tp\Big)^2
\int_0^1du { (1-u)^{4-D} \over \Big[V^2\Big(u\tp\Big) \Big]^{D/2}}
\label{parts}
\eea
and by considering that, for $D=4+2 \epsilon$, one can write
\be
\int_0^1du { (1-u)^{4-D} \over \Big[V^2\Big(u\tp\Big) \Big]^{D/2}}=
{1 \over V^2\Big(\tp \Big)} \Big\{F_2\Big( \tp \Big)-\epsilon 
\Big[ 2F_4\Big(\tp\Big)+F_6\Big(\tp\Big)\Big]\Big\}+{\cal O}(\epsilon^2)\;, 
\nonumber 
\ee
\noindent with the functions $F_2(z),F_4(z)$ and 
$F_6(z)$ reported in the Appendix.
Moreover, the first integral on the r.h.s. of 
eq. (\ref{parts}) can be evaluated by performing
one more integration by parts and making
use of the identity
\be
D\int_0^1 du {\ln^n(1-u) \Big(u\tp+y \Big) \over \Big[V^2\Big(u\tp\Big) \Big]
^{D/2+1}}=n\ttp
\int_0^1 du {\ln^{n-1}(1-u) \over 1-u} \Bigg[{1 \over \Big[V^2\Big(\tp\Big)
\Big]^{D/2}}-{1 \over \Big[V^2\Big(u\tp\Big)\Big]^{D/2}}\Bigg] \; , 
\label{log} 
\ee
with the result
\bea
\int_0^1 &du& { (1-u)^{3-D} \over \Big[V^2\Big(u\tp\Big) \Big]^{D/2}} =
{-1 \over \Big[V^2\Big(\ttp\Big)\Big]^2}
\cdot \Bigg\{{1\over 2 \epsilon}
- \Bigg[F_9\Big(\tp\Big)+{1 \over 2}\ln{V^2\Big(\tp\Big)}\Bigg]
\nonumber \\
&+& \epsilon\; \Bigg[{1\over 4}\ln^2{V^2\Big(\tp\Big)}+
F_9\Big(\tp\Big)\ln{V^2\Big(\tp\Big)}+F_8\Big(\tp\Big)+2F_7\Big(\tp\Big)\Bigg]
 +{\cal O}(\epsilon) \Bigg\}  \label{3-d}
\eea
(the expressions for the functions $F_7(z), F_8(z)$ and
$F_9(z)$ can be found in the Appendix).
The resulting expression for $D_5^{(1)}$
(the corresponding contribution to $D_6$ can obtained analogously) appears
rather simple, in spite of the involved expressions of the intermediate
formulae; as a matter of fact, one has  
\be
\hat D^{(1)}_5={Z_\mu \over 4}\Bigg\{{1 \over \epsilon}\; \ttp\Big(y+\tp\Big)-
2(y^2-1)F_1\Big(\tp\Big)-\ttp\Big(y+\tp\Big)\ln{V^2\Big(\tp\Big)} 
+{\cal O}(\epsilon) \Bigg\}\;.
\label{d51exp}
\ee

The contribution  $D_5^{(2)}$,
\bea 
D_5^{(2)}&=&8iN_cg_s^2C_F Tr\Bigg[\gamma_5 {1+ \spur{v} \over 2} \gamma_\mu
\gamma_5 {1+ \spur{v}^\prime \over 2} {1 \over 2}[\gamma^\alpha, \gamma^\beta]
\Bigg] \label{d52} \\
\int {d^D s \over (2\pi)^D} {d^D t \over (2\pi)^D}&& 
{s_\alpha t_\beta \over (\omega+2 v \cdot s)(\omega+2 v \cdot t)
(\omega^\prime+2 v^\prime \cdot s)s^2 t^2(s-t)^2}
\;, \nonumber
\eea
after double-Borel transform, can be written is terms of a triple integral:
\be
\hat D_5^{(2)}=8iN_cg_s^2C_F(1+y)(v-v^\prime)_\mu { \Gamma(D-1) \over (4\pi)^D}
\int_0^{1/\tau} d\lambda_1 {\lambda_1 \over \tau^\prime} \int_0^1 dz_1 \int_0^1 
dz_2 { z_1 (z_1^2 \bar z_1 z_2 \bar z_2)^{D/2-1} \over (k^2)^{D-1}} 
\label{d52a}
\ee
where: $\bar z_i=1-z_i$; $k^2=z_1(q^2\bar z_2+p^2z_2)-
z_1^2(q\bar z_2+pz_2)^2$, with 
$p=\lambda_1v$ and $q={v \over \tau} +{v^\prime \over \tau^\prime}$. 
The integral in (\ref{d52a}) can be simplified by
changing the variables: 
\be
\lambda_1={\lambda \over \tau}\hskip 2 cm z_1={\lambda u_2-2\Big(1-{u_1\over 2 
\lambda}\Big) \over u_1 u_2 -1} \hskip 2 cm z_2={ {u_1 \over \lambda}-1 \over 
\lambda u_2-2\Big(1-{u_1\over 2\lambda}\Big) }  
\ee
and by integrating over $\lambda$, obtaining:
\bea
\hat D_5^{(2)}&=& {Z_\mu \over 2} \ttp { (1+y) \over D-3} \Big[ V^2\Big(
\ttp\Big) \Big]^{D/2}  \nonumber \\
 \Bigg\{& -&{\Gamma(D-1) \over \Gamma\Big({D \over 
2}\Big) \Gamma\Big({D \over 2}-1\Big) } \int_1^\infty du_1 \int_1^\infty du_2 
{(u_1 u_2 -1)^{D/2-2} \over \Big[ u_2 V^2\Big(\ttp
\Big)+u_1-2\Big(1+y\ttp\Big) \Big]^{D-1} }\label{d52hat} \\
&-&\Big( \tp \Big)^D { 1 \over \Big[ V^2\Big(\ttp\Big) \Big]^{D/2-1}} 
\int_0^1 du 
{ 1-u \over \Big[ V^2\Big(u\tp\Big) \Big]^{D/2}} +\Big(\tp \Big)^D 
\int_0^1 du
{ (1-u)^{3-D} \over \Big[ V^2\Big(u\tp\Big) \Big]^{D/2} } \Bigg\}
\;. \nonumber
\eea
The corresponding contribution to $D_6$ can be obtained 
analogously. 
Also in this case it is important to notice the dependence on the
ratio of Borel parameters $\tau$ and $\tau^\prime$.

When we  expand $\hat D_5^{(2)}$ and
$\hat D_6^{(2)}$ in $\epsilon$ we obtain a quite simple result, namely:
\bea
\hat D_5^{(2)}&=&-{Z_\mu \over 2}(1+y)\ttp \Big[{1 \over 2 \epsilon}-{1 \over 
2}\ln{V^2\Big(\tp\Big)}-\Big(y+\tp\Big)\tp F_1\Big(\tp\Big) 
+{\cal O}(\epsilon) \Big]\;, 
\label{d52exp}
\\
\hat D_6^{(2)}&=&-{Z_\mu \over 2}(1+y)\Big[-{1 \over 2 \epsilon}
+{1 \over2}\ln{V^2\Big(\ttp\Big)}-\ln\Big(\tp\Big) +\ttp\Big(y+\ttp\Big)F_1\Big
(\ttp\Big) +{\cal O}(\epsilon) \Big]\;,
\label{d62exp}
\eea
\noindent and, for the sum $\sum_{i=1,2}(D_5^{(i)}+D_6^{(i)})$: 
\bea
&\sum_{i=1,2}&(D_5^{(i)}+D_6^{(i)})={Z_\mu \over 2} \Bigg\{  
\Big(1-\ttp\Big){1 \over \epsilon}\nonumber \\
&-&\Big(1-\ttp\Big)\ln{V^2\Big(\ttp\Big)}+
(1+y)\Big(1+\ttp\Big)\Big[\tp F_1\Big(\tp\Big)-\ttp F_1\Big(\ttp\Big)\Big]
+ {\cal O}(\epsilon) 
\Bigg\}\;. \label{d5612}
\eea

Let us finally consider the third contribution to $D_5$:
\be
D_5^{(3)}=8 i N_c g_s^2C_F Tr\Bigg[\gamma_5 {1+ \spur{v} \over 2} \gamma_\mu
\gamma_5 {1+ \spur{v}^\prime \over 2} \gamma^\alpha
\Bigg]\cdot I_\alpha \;,\label{d53} 
\ee
\noindent where
\be
I_\alpha=
\int {d^D s \over (2\pi)^D} {d^D t \over (2\pi)^D}
{s_\alpha \; 2 v \cdot t \over (\omega+2 v \cdot s)(\omega+2 v \cdot t)
(\omega^\prime+2 v^\prime \cdot s)\;s^2 \; t^2 \;(s-t)^2}
\;\;\; .\label{d53int}
\ee
\noindent
The integral  (\ref{d53int}) can be related to simpler ones 
by using the identity \cite{tkachov,neubert93}
\bea
-(D-4)I_\alpha&=& \int{d^D s \over (2\pi)^D} {d^D t \over (2\pi)^D} 
{s_\alpha \over 
(\omega+2 v \cdot s)
(\omega^\prime+2 v^\prime \cdot s)\; s^2 \; t^2\;(s-t)^2}\nonumber \\
&+& \int{d^D s \over (2\pi)^D} {d^D t \over (2\pi)^D} 
{ 2 v \cdot t \;s_\alpha \over 
(\omega+2 v \cdot s)(\omega+2 v \cdot t)
(\omega^\prime+2 v^\prime \cdot s)\; t^4\;(s-t)^2}\nonumber \\
&-& \int{d^D s \over (2\pi)^D} {d^D t \over (2\pi)^D} 
{ 2 v \cdot t \;s_\alpha \over 
(\omega+2 v \cdot s)(\omega+2 v \cdot t)
(\omega^\prime+2 v^\prime \cdot s) \;s^2\; t^4}\nonumber \\
&-& \int{d^D s \over (2\pi)^D} {d^D t \over (2\pi)^D} 
{ \omega \;s_\alpha \over 
(\omega+2 v \cdot t)^2
(\omega^\prime+2 v^\prime \cdot s) \;s^2 \;t^2 \;(s-t)^2}=\nonumber \\
&=&J_\alpha^{(1)}+J_\alpha^{(2)}+J_\alpha^{(3)}+J_\alpha^{(4)} \;. \label{ji}
\eea
\noindent After  Borel transformation, 
the results for $J_\alpha^{(i)}$, $i=1,2,3$ are given by
\bea
\hat J_\alpha^{(1)}&=&{4 \over (D-2)(D-4)}\;{\tau^{2D-5} \over (4\pi)^D} \Gamma 
\Big({D \over 2}-1 \Big)\Gamma\Big({D \over 2}\Big)
\Big(v+\ttp v^\prime\Big)_\alpha{ 1 \over \Big[V^2\Big(\ttp\Big)\Big]^{D-2}}
\label{j1}\\
\hat J_\alpha^{(2)}&=&-{2 \over \Big[V^2\Big(\ttp\Big)\Big]^{D/2-1}}\;
{\tau^{2D-5} \over (4\pi)^D} \Gamma
\Big({D \over 2}-1 \Big)\Gamma\Big({D \over 2}\Big)\label{j2} \\ 
&\Bigg\{& \Bigg[{ \Big(
1+y\ttp \Big) \over \Big[V^2\Big(\ttp\Big)\Big]} \Big(v +\ttp v^\prime 
\Big)_\alpha-{v_\alpha \over D-2} \Bigg] \Big(\tp\Big)^{D-2} \int_0^1 du\;
{1 \over \Big[V^2\Big(u\tp\Big)\Big]^{D/2-1}} \nonumber \\
&+& \Big(1+y\ttp \Big) \Big(\tp\Big)^D v_\alpha
\int_0^1 du\;{u \over \Big[V^2\Big(u\tp\Big)\Big]^{D/2}}+
\Big(1+y\ttp \Big) \Big(\tp\Big)^{D-1} v^\prime_\alpha
\int_0^1 du\;{1 \over \Big[V^2\Big(u\tp\Big)\Big]^{D/2}} \Bigg\} \nonumber\\
\hat J_\alpha^{(3)}&=&2\;{\tau^{2D-5} \over (4\pi)^D} \Gamma
\Big({D \over 2}-1 \Big)\Gamma\Big({D \over 2}\Big) \Bigg\{
\Big(\tp\Big)^D v_\alpha \int_0^1 du\;{u (1-u)^{3-D} \over 
\Big[V^2\Big(u\tp\Big)\Big]^{D/2}}   \label{j3}\\
&+& 
\Big(\tp\Big)^{D-1} v^\prime_\alpha \int_0^1 du\;{ (1-u)^{3-D} \over 
\Big[V^2\Big(u\tp\Big)\Big]^{D/2}} \Bigg\} \;.\nonumber
\eea
As for $J_\alpha^{(4)}$, it  can be obtained as the result of a differential 
equation introduced in \cite{kotikov,neubert93}, which gives
\bea
\hat J_\alpha^{(4)}&=&-{2 \over D-2}\;{\tau^{2D-5} \over (4\pi)^D} \Gamma
\Big({D \over 2}-1 \Big)\Gamma\Big({D \over 2}\Big) \cdot \label{j4}\\ 
&\cdot& 
\Bigg\{ \Big(\ttp\Big)^{3-D}{v^\prime_\alpha \over \Big[V^2\Big(\ttp\Big)
\Big]^{D/2-1}} -{D-2 \over D-4}\Big[1-\Big(\ttp\Big)^{4-D}\Big]{\Big(v +\ttp 
v^\prime \Big)_\alpha \over \Big[V^2\Big(\ttp\Big)\Big]^{D/2}} 
\nonumber \\
&&\;- (D-4)\Bigg[ \Big(\tp \Big)^{2D-5} v^\prime_\alpha \int_0^1du {u^{D-4} 
\over \Big[V^2\Big(u\tp\Big)\Big]^{D/2-1}}\nonumber \\
&\;&\hskip 2 cm
-{D-2 \over D-4} \Big( \tp 
\Big)^{2D-5}\int_0^1du {u^{D-4}
\over \Big[V^2\Big(u\tp\Big)\Big]^{D/2}}\Big[\Big(u \tp \Big)^{4-D}-1\Big] 
\Big(v^\prime +u \tp v \Big)_\alpha \Bigg]\Bigg\} \;\;\;. \nonumber
\eea 
Summing up the four contributions one can write
\be
\hat D_5^{(3)}=-{Z_\mu \over D-4}[\hat D_5^{(3.1)}+\hat D_5^{(3.2)}+
\hat D_5^{(3.3)}+\hat D_5^{(3.4)}] \;, \label{d53sum}
\ee
\noindent where
\bea
\hat D_5^{(3.1)}&=&-{2 \over (D-2)(D-4)}{1 \over \Big[V^2\Big(\ttp\Big)
\Big]^{D/2-2}}\Big(1-\ttp \Big)\;, \label{d531}\\
\hat D_5^{(3.2)}&=&-\Big[V^2\Big(\ttp\Big) \Big]\Bigg\{ \Bigg[-\Big(1-\ttp\Big)
{\Big(1+y\ttp \Big) \over \Big[V^2\Big(\ttp\Big)\Big]}+{1 \over D-2} \Bigg]
\Big(\tp \Big)^{D-2} \int_0^1du { 1 \over \Big[V^2\Big(u\tp\Big)
\Big]^{D/2-1}} \nonumber \\
&&\hskip 2.5 cm +\Big(1+y\ttp \Big)\Big(\tp \Big)^{D-1}
\int_0^1du { \Big(1-u \tp \Big) \over \Big[V^2\Big(u\tp\Big)\Big]^{D/2}} 
\Bigg\}\;, \label{d532}\\
\hat D_5^{(3.3)}&=&\Big(\tp \Big)^{D-1}\Big[V^2\Big(\ttp\Big) \Big]^{D/2}
\int_0^1du { (1-u)^{3-D} \over \Big[V^2\Big(u\tp\Big)\Big]^{D/2}}
\Big(1-u \tp \Big)\;, \label{d533}\\
\hat D_5^{(3.4)}&=&-{1 \over D-2}\Bigg\{\Big(\ttp \Big)^{3-D}
\Big[V^2\Big(\ttp\Big) \Big]+{D-2\over D-4}\Big[1-\Big(\ttp\Big)^{4-D}\Big] 
\Big(1-\ttp\Big) \label{d534} \\
&-&(D-4)\Big(\tp\Big)^{2D-5}\Big[V^2\Big(\ttp\Big)\Big]^{D/2}
\Bigg[\int_0^1du { u^{D-4} \over \Big[V^2\Big(u\tp\Big)\Big]^{D/2-1}}\nonumber 
\\
& -&{D-2\over 
D-4}\int_0^1du { u^{D-4} \over \Big[V^2\Big(u\tp\Big)
\Big]^{D/2}}\Big[\Big(u\tp\Big)^{4-D}-1\Big]\Big(1-u \tp \Big)\Bigg]\Bigg\}
 \nonumber \;.
\eea
The corresponding contribution to $D_6$ is obtained analogously.
It is worth noticing that some of the integrals in
$\hat D_5^{(3)}$ and  $\hat D_6^{(3)}$ require an expansion up to order 
$\epsilon^2$ to take care of the $D-4$ factor appearing on the l.h.s.
of (\ref{ji}). 

Despite the involved structure of the expressions for 
the various terms in $\hat D_5^{(3)}$ and $\hat D_6^{(3)}$, 
the result for the sum $\hat D_5^{(3)}+\hat D_6^{(3)}$ is rather simple:
\bea
\hat D_5^{(3)}+\hat D_6^{(3)}&=& Z_\mu \cdot \Bigg\{-{1 \over \epsilon} \Big(
1-\ttp \Big)
+(1+y)\Big(1+\ttp\Big)\Big[\ttp F_1\Big(\ttp\Big)- \tp F_1\Big(\tp\Big)\Big]
\nonumber \\
&+&\Big(1-\ttp\Big) \Big[4(y^2-1)F_1\Big(\ttp\Big)F_1\Big(\tp\Big)+{2 \over 3 }
\pi^2+2+2\ln{V^2\Big(\ttp\Big)} \Big]
+ {\cal O}(\epsilon) \Bigg\} \label{d563} \;,
\eea
\noindent 
where one can notice that spurious $\displaystyle{1 \over \epsilon^2}$
terms cancel out.

\subsection{ Final result}
The sum of the contributions of the diagrams $D_1-D_6$ gives the result:
\bea
\hat D =
\sum_{i=1}^6 \hat D_i&=&{ Z_\mu \over 2} \Big(1-\ttp\Big) 
\Bigg\{-{1 \over \epsilon} 
\Big[{5 \over 2}-yr(y) \Big]+ (1+y)\Big(1+\ttp\Big) G\Big(\ttp)
\nonumber \\
&+&\Bigg[y\Big[h\Big(y,\ttp\Big)+h\Big(y,\tp\Big)\Big]+yr(y)
\ln{V^2\Big(\tp\Big)}-y\sqrt{y^2-1}r^2(y)-\ln\Big(\tp\Big)^2\nonumber \\
&+&{1 \over 2}\ln{V^2\Big(\ttp\Big)}+
[4(y^2-1)-2y\sqrt{y^2-1}]F_1\Big(\ttp\Big)
F_1\Big(\tp\Big)
+{2 \over 3}\pi^2+6\Bigg] + {\cal O}(\epsilon) \Bigg\}
\nonumber \\
 \label{totale} 
\eea
where $G(x)=(x F_1(x)- {1\over x} F_1({1\over x}) )/(1-x)$.
The important point to notice is the structure of the 
$1/\epsilon$ singularity
in (\ref{totale}), which does not depend on the Borel parameters
$\tau, \tau^\prime$.

Eq.(\ref{totale}) represents, for all values of the Borel parameters 
$\tau$ and $\tau^\prime$ the ${\cal O}(\alpha_s)$ 
correction to the triangle diagram
representing the correlator (\ref{threep}). In particular, in the limit
$\tau \simeq \tau^\prime$ the expression can be simplified, giving
\bea
{\hat D}={\hat D}_0 {\alpha_s \over \pi} \Bigg\{& -&{1 \over {\hat \epsilon}} 
\Big[1-{\gamma(y)\over 2}\Big]+2 \gamma_E \Big[1-{\gamma(y)\over 2}\Big]
+2 \ln\Big({\mu \over \tau}\Big)\Big[1-{\gamma(y)\over 2}\Big] 
+ {4 \over 3}y h(y) \nonumber \\
&+& \Big[{2 \over 3} (y^2-1)-y\sqrt{y^2-1}\Big]r^2(y)+
\Big({2 
\over 3}y r(y)+{1 \over 3}\Big)\ln[2(1+y)]+{4 \over 9}\pi^2+{8 \over 3} \Bigg\}
 \label{d} 
\eea
where $\gamma_E$ is the Euler constant and 
${1 \over {\hat \epsilon}}={1 \over  \epsilon}+\gamma_E-\ln 4\pi$;
$\gamma(y)$ was defined in (\ref{andim}), and $h(y)=h(y,1)$. 
In the$ \overline{MS}$ subtraction scheme the ${1 \over {\hat 
\epsilon}}$ pole  cancels with the renormalization factors of the 
heavy-light and heavy-heavy quark currents in the correlator 
(\ref{correlatore}) \cite{hqet_1,renorm,broad91}:
${\cal Z}_{hl}^2=1-{\alpha_s \over \pi {\hat \epsilon}} \;, \hskip 1 cm 
{\cal Z}_{hh}=1+{\alpha_s \over2 \pi {\hat \epsilon}}\gamma(y)$,
so that the finite part represents the correction to the Borel-transformed
correlator we are looking for.

It is possible from eq.(\ref{d}) to determine  the 
spectral function $\rho^{pert}$ in (\ref{qcd}) at the order $\alpha_s$, 
which is required to perform the continuum subtraction
in the QCD sum rule analysis.
After changing the variables in 
(\ref{bor}) to $\sigma_\pm=\nu\pm\nu^\prime$, and 
integrating in $\sigma_-$, with integration limits 
$0 \le \sigma_+ \le +\infty$ and $-r\sigma_+ \le \sigma_- \le 
r\sigma_+$ ( $r=\sqrt{{y-1 \over y+1}}$),
one is left, for $\tau \simeq \tau^\prime$, with a function proportional to
$\sigma_+^3\Big[\rho_1(y)+\rho_2(y)\ln \Big({\mu \over \sigma_+}\Big) \Big]$
so that
\be
\hat D \propto \rho_1(y)+\Big[\ln \Big({\mu \over
\tau}\Big)+\gamma_E-{11 \over 6} \Big]\rho_2(y). \label{drho}
\ee
Comparing (\ref{drho}) with  (\ref{d}) we get, with $\hat D_0$ the $\alpha_s=0$ 
term (\ref{d0}):
\bea
\rho_1(y)&=&\hat D_0{\alpha_s \over \pi} \Bigg\{
 {4 \over 3}y h(y)+\Big[{2 \over 3} (y^2-1)-y\sqrt{y^2-1}\Big]r^2(y)+
\Big({2 \over 3}y r(y)+{1 \over 3}\Big)\ln[2(1+y)] \nonumber \\
&+&{4 \over 9}\pi^2+{8 \over 3}+{11 \over 3}\Big[1-{\gamma(y) \over 2}\Big]
 \Bigg\}\label{rho1} \\
\rho_2(y)&=&2 \hat D_0 {\alpha_s \over \pi} \Big[1-{\gamma(y) \over 2}\Big]
\; \label{rho2}
\eea
and therefore the spectral function $\rho^{pert}(\nu,\nu^\prime,y)$.

\section{Non-perturbative contributions and final sum rule}
Once the perturbative contribution to the sum rule has been computed, one has 
to consider non-perturbative power corrections, and, as anticipated, we include 
the vacuum condensates up to dimension five.

The lowest dimensional term in the expansion of
$\Pi^{np}$ in eq. (\ref{qcd}) is
the quark condensate $<{\bar q}q>$. At the tree level, the 
Borel transformed result for this contribution is simply given by
\be
\hat D_0^{<{\bar q}q>}=-i <{\bar q}q> (v-v^\prime)_\mu \;.\label{doqbq}
\ee
The ${\cal O}(\alpha_s)$ correction to (\ref{doqbq}) is 
computed from eight diagrams obtained from those in fig.2 replacing the light 
quark line by the quark condensate contribution to the relevant propagator.
The correction reads 
\be
\sum_{i=1}^8 D_i^{<\bar q q>}=- \hat D_0^{<\bar qq>}
\Big( {\alpha_s \over \pi} \Big) 
H(\tau,\tau^\prime)
\ee
where
\bea
H(\tau,\tau^\prime) &=& {2 \over 3} 
\Bigg\{ -2[1-yr(y)]\gamma_E-
[1-yr(y)]\Big(\ln \Big({\mu \over \tau}\Big)+ \Big({\mu \over \tau^\prime}\Big)
\Big) + \ln \Big(\tp\Big)[1+yr(y)]
\nonumber \\
&+&\ln V^2\Big(\ttp\Big) -5+(1+y)r(y) 
+ y \Bigg[ -h\Big(y, \ttp \Big)-h\Big(y, \tp \Big)-r(y)\ln  V^2\Big(\tp\Big)
\nonumber \\
&+&\sqrt{y^2-1}\; r^2(y)+2 \sqrt{y^2-1}\; F_1\Big(\ttp \Big) F_1 \Big(\tp \Big) 
\Bigg]  \Bigg\} \label{qbq}
\eea
with the notations previously defined. The calculation of
the dimension five contribution is straightforward.
Then, 
the final sum rule for $\tau_{1/2}$ is
\bea
2 \; \tau_{1/2}(y,\mu) F(\mu) F^+(\mu)&~&e^{-{2 \bar \Lambda \over \tau} -
{2 \bar \Lambda^+ \over \tau^\prime} } =
\int_D d \nu d \nu^\prime \rho^{pert}(\nu,\nu^\prime,y) 
e^{-{\nu \over \tau} - {\nu^\prime \over \tau^\prime} } \nonumber \\ 
&-& <\bar q q>(\mu) \Big(1 + {\alpha_s \over \pi} H(\tau,\tau^\prime) 
-{m_0^2\over 2} ({1\over 2 \tau^2} + {1 \over 2 \tau^{\prime 2}} +
{4 y+1 \over 3 \tau \tau^\prime} ) \Big) 
\label{fsumr}
\eea
where
\bea
\rho^{pert}(\nu,\nu^\prime,y)&=& {3 \over 16 \pi^2} 
{\nu^\prime-\nu \over (y-1) \sqrt{y^2-1} } \nonumber \\ 
&\Big[& 1 + {\alpha_s \over \pi} \Big( (2-\gamma(y)) \ln{\mu \over 
\nu^\prime+\nu} + {4 \pi^2 \over 9} + {19 \over 3} +c_p(y) \Big) \Big] 
\;\Theta(\nu^\prime-\nu^\prime_-) \;\Theta(\nu^\prime_+-\nu^\prime)
\;\;, \nonumber \\
\eea
and
\bea
c_p(y) &=& {4\over 3} y h(y) + 
\Big ({2\over 3} (y^2-1) - y \sqrt{y^2-1}\Big) r^2(y) 
\nonumber \\ 
&+& {\gamma(y)\over 2} \Big( - {11\over 3} + \ln(2(1+y)) \Big) + \ln(2(1+y))
\eea
($\nu^\prime_\pm=y_\pm \; \nu$). The integration domain $D$
is constrained by the conditions $\nu\le \nu_c$,
$\nu^\prime  \le \nu^\prime_c$.

Since the form factor $\tau_{1/2}$ is defined by the 
matrix elements of weak currents in the effective theory,
it depends on the subtraction scale $\mu$, and the sum rule (\ref{fsumr}) 
clearly reproduces this feature. 
As discussed in Sect.II, and in analogy with the case 
of the Isgur-Wise function $\xi$, it is possible to remove the 
scale-dependence by 
compensating it by the analogous $\mu$-dependence of the Wilson coefficients 
relating the $b \to c$ axial current in full QCD to the dimension 3 currents 
in HQET and by defining $\tau_{1/2}^{\rm ren}$ as in eq.(\ref{tauren}).
This is the function we shall consider in our numerical analysis.

\section{Numerical results}
The numerical analysis of the sum rule for $\tau_{1/2}$ can be
 carried out using 
the same input parameters adopted in Sect.IV for the determination of $F^+$.
In particular, we use the explicit expressions
of the two-point sum rules determining the leptonic constants
$F^+$ and $F$ that appear in the pole contribution of eq.(\ref{pole}).
We vary the threshold parameters in the ranges
$\nu_c=2-3$ GeV and $\nu^\prime_c=2.5-3.5$ GeV, obtaining 
an acceptable stability 
window, where the results do not appreciably depend 
on the Borel parameters, in the ranges around 
$\tau\simeq 1.5$ GeV and $\tau^\prime \simeq 2$ GeV, respectively.
The contribution of the nonperturbative term in the three-point correlator
represents a small fraction of the total contribution; on the other hand,
the $\alpha_s$ correction in the perturbative term is sizeable,
but it turns out to be partially compensated by the analogous 
correction in the leptonic constants $F$ and $F^+$. Notice that this is a 
remarkable result, not expected a priori since the normalization of the form 
factor, for example at zero recoil, is not fixed by symmetry arguments.
The perturbative corrections, however, do not equally affect the form factor
for all values of the variable $y$, but they are sensibly $y-$dependent,
with the effect of increasing the slope of $\tau_{1/2}$ with respect to the 
case where they are omitted. 

The results for $\tau_{1/2}^{ren}(y)$ are shown
in fig.3, where the curves refer to various choices for the continuum 
thresholds. The region limited by the curves essentially determines the 
theoretical accuracy allowed by the present calculation.

Considering the $y$ dependence, the limited range of values 
allowed by the mass difference between $D$ and $D_0$ permits the expansion near 
$y=1$: 
\be
\tau_{1/2}^{ren}(y)=\tau_{1/2}(1) 
\Big(1-\rho^2_{1/2} (y-1)+c_{1/2} (y-1)^2\Big) \;\;\;.
\ee
A two-parameter fit to fig.3, in terms of the normalization at zero recoil 
and the slope,  gives
$\tau_{1/2}(1)=0.31\pm0.06$ and $\rho^2_{1/2}=1.5\pm0.4$.
The inclusion of the quadratic term modifies the fit as follows:
\be
\tau_{1/2}(1)=0.35\pm0.08 \;\;, 
\hskip 1cm \rho^2_{1/2}=2.5\pm 1.0 \;\;, \hskip 1cm c_{1/2}=3\pm 3
\ee
which is the result we quote for our analysis.

The immediate application of this result concerns the prediction of the 
semileptonic $B$ decay rates to $D_0$ and $D_1^*$.
Using
$V_{cb}=3.9 \times 10^{-2}$ and 
$\tau(B)= 1.56 \times 10^{-12}$ sec, we obtain
\be
{\cal B}(B\to  D_0 \ell \bar \nu)=(5 \pm 3) \times 10^{-4} 
\hspace{1cm}
{\cal B}(B\to D^*_1 \ell \bar \nu)= (7 \pm 5) \times 10^{-4} \;\;\;.
\ee
This means that only a very small fraction of the semileptonic $B \to X_c$ 
decays  is represented by transitions into the
$s_\ell^P={1\over 2}^+$ charmed doublet. Although small, however,
one cannot exclude that such processes will be identified, mainly at dedicated 
$B$-facilities which will be running in the near future.
At present, the
measurements of semileptonic $B \to D^{**}$ decays
only provide data on the members of the $s_\ell^P={3\over 2}^+$ doublet
\cite{aleph95,cleo97}, since the doublet with $s_\ell^P={1\over 2}^+$
is not distinguished from the non-resonant charmed background.
In particular, in \cite{aleph95} the $B$ semileptonic
branching fraction to the final 
states $D\pi$ and $D^*\pi$ of $(20\pm 5)\;10^{-2}$ is reported.

\section{Conclusions}
We conclude observing that HQET has proven to be 
a powerful tool to 
handle heavy quark physics. However, predictions derived in this framework 
should always be supported by the computation of $1/m_Q$ as well as radiative 
corrections. The role of both depend on the specific situation one is 
facing with. For example, they turn out to be important for the $B$ meson 
leptonic constant, while they are moderate for the Isgur-Wise  function, as 
derived in \cite{neubert93}. We have presented here the case of the universal 
form factor $\tau_{1/2}(y)$ describing $B$ semileptonic transitions to the 
excited $J^P=(0^+,1^+)$ charmed states, using QCD sum rules in the framework of 
HQET. As already shown in \cite{neubert93}, the computation of loop integrals 
results to be greatly simplified within  HQET. 
The task of 
computing perturbative corrections to $\tau_{1/2}(y)$ is justified by manifold 
interesting phenomenological features of orbitally excited states as well as by 
the many theoretical interests already mentioned. We have obtained 
a situation similar to the case of the Isgur-Wise function, namely radiative 
corrections are quite under control for $\tau_{1/2}(y)$, while they affect 
considerably  the 
value of the leptonic constant $F^+$ of the $s_\ell=1/2$ doublet.

\vspace{1cm}
\acknowledgments{ We thank G. Nardulli for interesting discussions.}

\newpage
\appendix
\section{Parametric Integrals}
The calculation of the two-loop 
diagrams relevant for the form factor $\tau_{1/2}$ 
essentially follows the analogous calculation for the Isgur-Wise function 
\cite{neubert93}, with the main difference represented by the need of keeping 
different Borel parameters, due to the non-symmetric nature of the problem at 
hand. We only recall here that, in momentum space,
the Feynman rules  of HQET \cite{hqet_2,falk90,reviews}
can be reduced to the heavy quark propagator
${ \spur{v} +1 \over 2} {i \over v\cdot k}$, and to the 
heavy-quark-gluon vertex $i g_s v_\mu {\lambda^a \over 2}$;
$N_c$ is the number of colours and
$Tr[{\lambda^a \over 2}{\lambda^a \over 2}]=C_F={N_c^2-1 \over 2 N_c}$.

The calculation of the loop integrals is performed 
in $D=4+2 \epsilon$ Euclidean space-time dimensions. The main ingredients
are the representations of the propagators of the massless quark and of the
heavy quark: 
\bea
{1 \over (s_E^2)^a}&=&{\Gamma\Big({D \over 2}-a \Big) \over \pi^{D/2} \Gamma(a)}
\int d^D x {e^{2is_E\cdot x} \over (x^2)^{D/2-a}}
\label{fourier} \\
{1 \over (\omega+2 iv_E \cdot s_E)^\alpha}&=&{(-1)^\alpha \over \Gamma(\alpha)}
\int_0^\infty d\lambda \lambda^{\alpha-1} e^{\lambda(\omega +2iv_E\cdot s_E)}
\label{hprop}
\eea
\noindent 
($v_E, s_E$ obtained from the four-vectors $v,s$ by a Wick rotation); 
in particular, 
(\ref{hprop}) is useful for the computation of the integrals after 
Borel transformation, since
\be
{\hat B}^{(\omega)}_\tau e^{\lambda\omega}=\delta(\lambda-\tau^{-1})
\;\;\; . \label{borexp}
\ee
\noindent
The master integrals needed in the evaluation of the loop integrals 
can be found in \cite{neubert92a,broad91}.
Here we report a number of parametric integrals useful for
the calculation of the Borel-transformed expressions $\hat D_1-\hat D_6$:

\bea
F_1(z)&=&\int_0^1 {du \over V^2(uz)}=-{1 \over 2z\sqrt{y^2-1}} \;
\ln\Big[{y_+ +z \over y_+^2(y_-+z)}\Big] \;,\label{f1}
\nonumber \\
F_2(z)&=&V^2(z)\int_0^1{du \over \Big[V^2(uz)\Big]^2}={1 \over 2(y^2-1)} 
[y(y+z)+y^2-1-V^2(z)F_1(z)] \;,\label{f2}
\nonumber \\
F_4(z)&=&V^2(z)\int_0^1du{\ln(1-u) \over \Big[V^2(uz)\Big]^2}=
-{1 \over 2}F_1(z)-\ln{V^2(z)}\Big(1+{y \over z}\Big){1 \over 4 (y^2-1)} 
 \label{f4} \nonumber \\
&+&V^2(z){1 \over 4 z (y^2-1)}\Big[{\cal L}(y,z)+{1 \over z}
\ln{V^2(z)}F_1\Big({1 
\over z}\Big)-z\ln{V^2\Big({1 \over z}\Big)}F_1(z)\Big]\;,
\nonumber
\\
F_5(z)&=&\int_0^1du{\ln{V^2(uz)} \over V^2(uz)}\label{f5} \nonumber\\
&=&-{1 \over z}h(y,z)-
{1 \over z^2}F_1\Big({1 \over z}\Big)\ln{V^2(z)}+\sqrt{y^2-1}F_1(z)
\Big[r(y)+{1 \over z}F_1\Big({1\over z}\Big) \Big]\;, \nonumber 
\\
F_6(z)&=&V^2(z)\int_0^1du{\ln{V^2(uz)} \over \Big[V^2(uz)\Big]^2}
={V^2(z) \over 2(y^2-1)}\Bigg\{F_1(z)-F_5(z)+{y \over z}-{\Big({y \over 
z}+1\Big) \over V^2(z)} [1+\ln{V^2(z)}] \Bigg\} \;,\label{f6}
\nonumber \\
F_7(z)&=&\int_0^1du{\ln(1-u) \over 1-u}\Bigg[{[V^2(z)]^2 \over [V^2(uz)]^2}-1 
\Bigg] 
=(y+z)zF_4(z)\Big[1-{2(y^2-1)\over V^2(z)}\Big]\label{f7}  \\
&-&{z \over 2}(y+z)F_1(z)
\Big[1+{2(y^2-1)\over V^2(z)}\Big]-{1 \over 4}\ln{V^2(z)}
\Big[3+{2(y^2-1)\over V^2(z)}\Big]
+{1 \over 4}\ln{V^2(z)}\ln{V^2\Big({1 \over z}\Big)}\nonumber \\
&-&(y^2-1)F_1(z)
F_1\Big({1 \over z}\Big)-{\pi^2 \over 6}+{1 \over 2}{\cal L}_1(y,z)\;,
\nonumber
\\
F_8(z)&=&[V^2(z)]^2\int_0^1{du \over 1-u} {\ln{V^2(uz)}-\ln{V^2(z)} \over 
[V^2(uz)]^2}=-{\cal L}_1\Big(y,{1\over z}\Big)-{1 \over 4}\big[\ln{V^2(z)}
\big]^2
\label{f8} \nonumber \\
&-&(y+z)\ln{V^2(z)}zF_1(z)-{1 \over 2}+{1 \over 2}V^2(z)
\big[1-\ln{V^2(z)}\big]\nonumber \\
&+&z(y+z)\big[F_5(z)+F_6(z)-\ln{V^2(z)}F_2(z
)\big] \;, \nonumber
\\
F_9(z)&=&\int_0^1{du \over 1-u}\Bigg[{[V^2(z)]^2 \over [V^2(uz)]^2}-
1 \Bigg] 
=z(y+z)\big[F_1(z)+F_2(z)\big]+{\big[\ln{V^2(z)}+V^2(z)-1\big] \over 2}
\;.\label{f9} \nonumber
\eea
\noindent 
In $F_1-F_8$ the combinations have been introduced:
\bea
h(y,z)&=&{ 1\over \sqrt{y^2-1}}\Bigg[-L_2(1-y_-^2)+L_2\Big({y_+(1-y_-^2) \over
y_++z}\Big)\Bigg] \;,\nonumber\\
{\cal L}(y,z)&=&{ 1\over \sqrt{y^2-1}}\Bigg[L_2\Big({1 \over 1+y_-z}\Big)-
L_2\Big({1 \over 1+y_+z}\Big)\Bigg]\;,  \\
{\cal L}_1(y,z)&=&L_2\Big({1 \over 1+y_-z}\Big)+
L_2\Big({1 \over 1+y_+z}\Big) \nonumber \;,
\eea
\noindent 
where $L_2(x)$ is the dilogarithm.
The integrals $F_1(z)-F_9(z)$ coincide with those reported in 
\cite{neubert93} for $z=1$.
Finally, a useful identity needed in the
calculation of $D_5^{(3)}$, $D_6^{(3)}$ is
\be
{\cal L}\Big(y,\ttp\Big)-{\cal L}\Big(y,\tp\Big)=-\tp F_1\Big(\tp\Big) 
\ln{V^2\Big(\ttp \Big)}+\ttp F_1\Big(\ttp\Big)
\ln{V^2\Big(\tp \Big)}\;.
\ee

\newpage
\hskip 3 cm {\bf FIGURE CAPTIONS}
\vskip 1 cm
{\bf Fig. 1} \par
\noindent
Binding energy parameter $\bar \Lambda^+$ and leptonic constant $F^+_{ren}$
of the doublet $s_\ell^P={1 \over 2}^+$,
from the QCD sum rule analysis of the correlator eq.(\ref{corr2pt}). 
The curves refer to three choices of the threshold 
parameter $\nu^\prime_c$: 
$\nu^\prime_c=2.5$ GeV (continuous line),
$\nu^\prime_c=3.0$ GeV (dashed line),
$\nu^\prime_c=3.5$ GeV (dotted line).

\vskip 1 cm
{\bf Fig. 2} \par
\noindent
Two-loop diagrams relevant for the calculation of ${\cal O} (\alpha_s)$
corrections to the perturbative part of the QCD sum rule for the
form factor $\tau_{1/2}$. The heavy lines represent the heavy quark propagators 
in HQET.

\vskip 1 cm
{\bf Fig. 3} \par
\noindent
The universal form factor $\tau_{1/2}^{ren}(y)$.
The curves refer to choices of the threshold parameters:
$\nu_c=2.0$ GeV,  $\nu^\prime_c=2.5$ GeV (continuous line),
$\nu_c=2.5$ GeV,  $\nu^\prime_c=3.0$ GeV (dashed line),
$\nu_c=3.0$ GeV,  $\nu^\prime_c=3.5$ GeV (dotted  line).

\newpage

\begin{figure}
\begin{center}
\mbox{%
\psfig{file=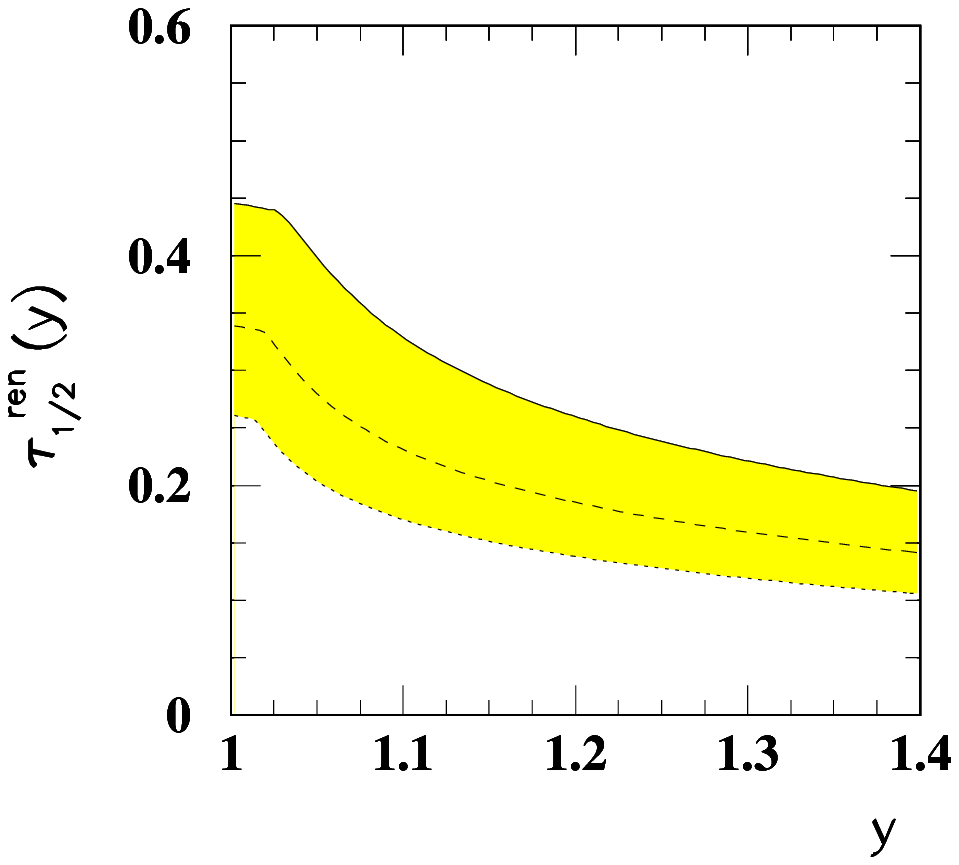,width=12cm}}
\end{center}
\caption{ } \label{fig_tau12}
\end{figure}
               
\end{document}